\documentclass[11pt, a4paper]{article}
\usepackage[top=1in, bottom=1in, left=1.25in, right=1.25in]{geometry}
\usepackage{booktabs}
\usepackage{multirow}
\usepackage{setspace}
\usepackage{natbib}
\usepackage{authblk}
\usepackage{graphicx}
\usepackage{subfigure} 
\usepackage{amsmath}
\usepackage{amsthm}
\usepackage{mathtools}

\DeclarePairedDelimiterX\Set[2]{\{}{\}}{\mspace{2mu}{#1}\;\delimsize|\;{#2}\mspace{2mu}}
\mathtoolsset{showonlyrefs}
\usepackage{amsfonts}
\usepackage{algorithm}
\usepackage{ascmac}
\usepackage{amssymb}
\usepackage{bm}
\usepackage{bbm}
\usepackage[table]{xcolor}
\usepackage{enumitem}
\usepackage{mathrsfs}
\usepackage[ 
	colorlinks = true, 
	anchorcolor=black,
	linkcolor = blue,
	citecolor = blue, 
	filecolor  = blue, 
	urlcolor   = cyan
]{hyperref}
\usepackage{comment}

\theoremstyle{plain}
\newtheorem{thm}{Theorem}

\theoremstyle{definition}

\theoremstyle{remark}
\newtheorem*{rem}{Remark}

\newcommand{\ep}{\varepsilon}

\newcommand{\E}{\mathbb{E}}

\newcommand{\mbb}[1]{\mathbb{#1}}

\newcommand{\mcal}[1]{\mathcal{#1}}

\def\dd{{\rm d}}

\title{\bf Divergence-based Robust Generalised Bayesian Inference for Directional Data via von Mises-Fisher models}

\author[1,2]{Tomoyuki Nakagawa}
\author[3]{Yasuhito Tsuruta}
\author[4]{Sho Kazari}
\author[4]{Kouji Tahata}

\affil[1]{School of Data Science, Meisei University}
\affil[2]{RIKEN Center for Brain Science}
\affil[3]{School of Business Administration, Meiji University}
\affil[4]{Department of Information Sciences, Tokyo University of Science}

\date{Last update: \today}

\begin{document}
\maketitle
\begin{abstract}
   This paper focusses on robust estimation of location and concentration parameters of the von Mises-Fisher distribution in the Bayesian framework. 
   The von Mises-Fisher (or Langevin) distribution has played a central role in directional statistics.
    Directional data have been investigated for many decades, and more recently, they have gained increasing attention in diverse areas such as bioinformatics and text data analysis.
    Although outliers can significantly affect the estimation results even for directional data, the treatment of outliers remains an unresolved and challenging problem. 
    In the frequentist framework, numerous studies have developed robust estimation methods for directional data with outliers, but, in contrast, only a few robust estimation methods have been proposed in the Bayesian framework. 
    In this paper, we propose Bayesian inference based on the density power divergence and the $\gamma$-divergence and establish their asymptotic properties and robustness.
    In addition, the Bayesian approach naturally provides a way to assess estimation uncertainty through the posterior distribution, which is particularly useful for small samples. 
    Furthermore, to carry out the posterior computation, we develop the posterior computation algorithm based on the weighted Bayesian bootstrap for estimating parameters.
    The effectiveness of the proposed methods is demonstrated through simulation studies. Using two real datasets, we further show that the proposed method provides reliable and robust estimation even in the presence of outliers or data contamination. 

\end{abstract}

\noindent{{\bf Keywords}: Directional data; Robust divergence; Generalised Bayesian Inference; von Mises-Fisher distribution}

\medskip

\noindent{{\bf Mathematics Subject Classification}: Primary 62H11; Secondary 62F15}
\section{Introduction}
\label{sec:intro}
Directional data that take values on the $(p-1)$-dimensional unit sphere 
\[\mcal{S}_{p} = \Set*{\bm{x} \in \mbb{R}^{p}}{ \|\bm{x}\| = 1}\]
arise in various scientific fields, where $\|\cdot\|$ is Euclidean norm. 
For example, wind direction, animal orientation, and wildfire orientation are distributed on the surface of $\mcal{S}_{2}$ \citetext{\citealp[Chapter 1]{mardia2000directional}; \citealp{garcia2014test}}. Typical examples for $p=3$ are the directions of paleomagnetism in rock and from the earth to the stars \citep[Chapter 9]{mardia2000directional}. Text data and Genome sequence representations are observations on a high-dimensional unit sphere \citep{banerjee2005clustering}. 
Consequently, the analysis of directional data has been attracting increasing attention across a wide range of fields.
For the analysis of spherical data, several probability distributions have been proposed in the literature \citep[e.g., ][so on]{mardia2000directional,pewsey2021recent}. 
Among these distributions, a model that has played a central role is the von Mises-Fisher (or Langevin) distribution. 
Its density is given by 
\begin{align*}
f(\bm{x} \mid \bm{\mu}, \kappa) = \frac{\kappa^{(p -2)/2}}{(2\pi)^{p/2} I_{(p-2)/2}(\kappa)}\exp\{\kappa \bm{\mu}^{\top}\bm{x}\}, \quad  \bm{x} \in \mcal{S}_{p}, 
\end{align*}
where $\bm{\mu}$ is the location parameter and $\kappa$ is the concentration parameter, with $\bm{\mu} \in \mcal{S}_{p}$, $\kappa \geq 0$. 
Here, $I_q(\cdot)$ denotes the modified Bessel function of the first kind and order $q$. 
Estimation of the location and concentration parameters of the von Mises-Fisher distribution has long been a central topic in directional statistics \citep[e.g.][]{mardia2000directional, pewsey2021recent}. 

In contrast to Euclidean space, the sample space for modelling directional variables is bounded.
Furthermore, the parameter space is often bounded as well; for instance, the location parameter of the von Mises-Fisher distribution lies on the unit sphere.
These characteristics pose fundamental challenges in defining outliers for directional data and in assessing the robustness of estimators. 
Although a large body of literature has been devoted to robust estimation methods for models with observations in $\mbb{R}^p$ \citep[see, for example, ][]{huber2009robust, hampel1986robust, maronna2019robust}, comparatively little attention has been paid to robust estimation for models defined on a bounded sample space. 
Most existing studies have focused on robust estimation of either the location or the concentration parameter of the von Mises-Fisher distribution.
However, comparatively few works have addressed the simultaneous estimation of both parameters.
\citet{lenth1981robust} briefly described a numerical algorithm for jointly estimating the two parameters in the circular case.
\citet{ko1993robust} proposed a method that combines two different approaches originally developed for estimating each parameter separately.
\citet{laha2012sb} introduced another robust estimator for the concentration parameter, while \citet{agostinelli2007robust} proposed two alternative methods for the joint estimation of both parameters in the circular case.
\cite{kato2016robust} proposed the robust divergence-based approach for the joint estimation of both parameters in general-dimensional settings.
Robust estimators for the parameters of several other circular distributions were later investigated by \citet{laha2019sb}.

In the Bayesian context, the location and concentration parameters of the von Mises-Fisher distribution have also long been studied in the literature \citep[see, for example,][]{damien1999full, giummole2019objective, Hornik2013bayesian, mardia1976bayesian,  pal2022modified}. 
However, robust estimation for directional data has not been studied enough, and the existing literature in Bayesian inference is very limited. 
In Bayesian statistics, robustness against outliers is also an important issue \citep[see, e.g.,][among others]{berger1994overview,o2012bayesian}. Over the past few years, a number of robust Bayesian estimation methods have been proposed, including approaches based on heavy-tailed distributions to accommodate outliers \citep[e.g.,][]{desgagne2015robustness, gagnon2020new, hamura2022log}, divergence-based methods \citep[e.g.,][]{futami2018variational, ghosh2016robust, hooker2014bayesian,  jewson2018principles, momozaki2023robustness, nakagawa2020robust, nakagawa2021default}, and loss-based approaches \citep[e.g.,][among others]{Kawakami02012023, matsubara2022generalized}. 
Nevertheless, these studies have primarily focused on data in Euclidean space, and have not considered data defined on compact sets such as directional data. 
Motivated by this perspective, we focus on divergence-based generalised Bayesian inference for directional data and consider robust loss functions derived from the density power divergence and the $\gamma$-divergence.

This paper provides a robust Bayesian inference framework for directional data using divergences via von Mises-Fisher models.
In this study, we propose the generalised posterior distribution for directional data using the density power divergence and the $\gamma$-divergence and investigate their robustness. 
This approach facilitates a straightforward assessment of estimation uncertainty, particularly in scenarios characterized by a limited sample size. 
Additionally, we provide the posterior computation algorithm based on the Bayesian bootstrap \citep{lyddon2019general, newton1994approximate,newton2021weighted, rubin1981bayesian} to sample from the proposed posterior distribution. 
This makes it possible to approximately sample from the posterior distribution even when $p$ is large, thereby making Bayesian inference feasible.
This paper is organised as follows: In Section \ref{sec:VonMises}, we explain a robust estimation for the von Mises-Fisher models based on the density power divergence and the $\gamma$-divergence. 
In Section \ref{sec:GBI}, we propose a robust posterior distribution based on the density power divergence and the $\gamma$-divergence. We also discuss the method for computing robust posterior distributions through sampling. 
While sampling methods based on MCMC are typically employed, they were not applicable to the proposed posterior distributions due to specific application constraints. 
Therefore, posterior computations were carried out using the Bayesian Bootstrap method in this study.
Section \ref{sec:Robust} discusses robustness of the proposed estimators using robust posterior distributions in directional data. 
We derive the standardised influence function of the proposed estimators and compare them with ordinary Bayesian inference. 
In Section \ref{sec:sim}, we present the numerical performance of posterior means based on the density power divergence and the $\gamma$-divergence in simulation studies, and we see that these posterior means are hardly influenced even in the presence of outliers.
In Section \ref{sec:DA}, we apply our proposed estimation to the wind direction data and the gene expression data, illustrating its effectiveness in practical settings.

\section{Robust estimation for von Mises-Fisher models}
\label{sec:VonMises}

It is well known that the MLE is highly sensitive to outliers. 
This has led to the development of many robust estimation methods to address this issue \citep[see, for example, ][]{hampel1986robust, huber2009robust, maronna2019robust}. 
One well-known robust estimation method is the divergence approach. 
For the probability density functions $f$ and $g$,  
\cite{basu1998robust} proposed the density power divergence defined as $\mcal{D}_{\alpha}(g, f) = -d_{\alpha}(g, g) + d_{\alpha}(g, f)$ where $d_{\alpha}(g, f)$ is the cross entropy given as
\begin{align*}
d_{\alpha}(g, f) & = 
-\frac{1}{\alpha}\int g(\bm{x}) f(\bm{x})^{\alpha}\dd \bm{x} + \frac{1}{(1+ \alpha)}\int f(\bm{x})^{1+\alpha}\dd \bm{x}, \quad (\alpha>0). 
\end{align*} 
\cite{jones2001comparison} also proposed the $\gamma$-divergence defined as $\mcal{D}_{\gamma}(g, f) =-d_{\gamma}(g, g) + d_{\gamma}(g, f)$ where $d_{\alpha}(g, f)$ is the cross entropy given as
\begin{align*}
d_{\gamma}(g, f)& =  
-\frac{1}{\gamma}\log\int g(\bm{x})f(\bm{x})^{\gamma}\dd \bm{x} + \frac{1}{(1+ \gamma)}\log\int f(\bm{x})^{1+\gamma} \dd \bm{x}, \quad (\gamma>0). 
\end{align*}
Here, these are consistent with Kullback-Leibler (KL) divergence when $\alpha, \gamma = 0$. 
\cite{basu1998robust} and \cite{jones2001comparison} proposed estimators by minimising the empirical versions of these divergences and their estimators are known to hold strong robustness. 

In the directional data, \cite{kato2016robust} introduced these estimators for the von Mises-Fisher distribution and examined their robustness.
 Let $\bm{\xi} = \kappa \bm{\mu}$, then the von Mises-Fisher distribution forms a canonical exponential family on the unit sphere with respect to the natural parameter $\bm{\xi}$, whose probability density function $f_{\bm{\xi}}$ can be expressed as 
\begin{align}
    f_{\bm{\xi}}(\bm{x}) = f(\bm{x}\mid \bm{\xi}) = \frac{\|\bm{\xi}\|^{(p-2)/2}}{(2\pi)^{p/2}I_{(p-2)/2}(\|\bm{\xi}\|)} \exp{(\bm{\xi}^{\top}\bm{x})}, \quad \bm{x} \in \mcal{S}_p; \quad \bm{\xi} \in \mbb{R}^p.  
\end{align}
It is noted that $\bm{\xi}$ is a parameter in $\mbb{R}^p$, and the Euclidean norm of $\bm{\xi}$ represents the concentration parameter $\kappa$, while the standardised parameter $\bm{\xi}/\|\bm{\xi}\|$ represents the mean direction parameter $\bm{\mu}$. 
Under this parametrisation, the cross entropy of the density power divergence between the von Mises-Fisher distribution $f_{\bm{\xi}}$ and an arbitrary density $g$ on $\mcal{S}_p$ is given as follows 
    \begin{align}
    d_{\alpha}(g,f_{\bm{\xi}}) &=  -\frac{1}{\alpha}\frac{1}{ K_p(\bm{\xi})^{\alpha} }\int_{\mcal{S}_p}\exp{(\alpha \bm{\xi}^{\top} \bm{x})} g(\bm{x}) \dd \bm{x} +  \frac{1}{\alpha + 1}\frac{K_p((1 + \alpha)\bm{\xi})}{K_p(\bm{\xi})^{\alpha + 1}}, 
    \end{align}
where $K_p(\bm{\xi}) = \{(2\pi)^{p/2}I_{(p-2)/2}(\|\bm{\xi}\|)\}/\|\bm{\xi}\|^{(p-2)/2}$.
Similarly, the cross entropy of the $\gamma$-divergence between $f_{\bm{\xi}}$ and an arbitrary density $g$ is given by
\begin{align*}
    d_{\gamma}(g,f_{\bm{\xi}}) = -\frac{1}{\gamma}\log\int_{\mcal{S}_p}\exp{(\gamma \bm{\xi}^{\top} \bm{x})} g(\bm{x}) \dd \bm{x} +  \frac{1}{\gamma + 1}\log K_p((1 + \gamma)\bm{\xi}).
\end{align*}

Let $\bm{x}_1, \ldots, \bm{x}_n$ be observations on $\mcal{S}_{p}$, and let $\bar{g}$ denote their empirical distribution, then the empirical cross entropies $d_{\alpha}(\bar{g}, f_{\bm{\xi}})$ and $d_{\gamma}(\bar{g}, f_{\bm{\xi}})$ are defined as follows:
\begin{align*}
d_{\alpha}(\bar{g},f_{\bm{\xi}}) &=  -\frac{1}{\alpha}\frac{1}{ K_p(\bm{\xi})^{\alpha} }\frac{1}{n}\sum_{i = 1}^n\exp{(\alpha \bm{\xi}^{\top} \bm{x}_i)} +  \frac{1}{\alpha + 1}\frac{K_p((1 + \alpha)\bm{\xi})}{K_p(\bm{\xi})^{\alpha + 1}},\\
    d_{\gamma}(\bar{g},f_{\bm{\xi}}) &= -\frac{1}{\gamma}\log\left(\frac{1}{n}\sum_{i = 1}^n\exp{(\gamma \bm{\xi}^{\top} \bm{x}_i)}\right) +  \frac{1}{\gamma + 1}\log K_p((1 + \gamma)\bm{\xi}).
\end{align*}
The estimators obtained by minimising these empirical cross entropies  $d_{\alpha}(\bar{g}, f_{\bm{\xi}})$ and $d_{\gamma}(\bar{g}, f_{\bm{\xi}})$ were proposed by \cite{kato2016robust}, who also investigated their robustness properties.
Although these point estimators hold strong robustness against outliers, it is not easy to evaluate the uncertainty in estimation, such as confidence intervals, due to the complexity of the associated estimating procedures.

\section{Robust Generalised Bayesian Inference via von Mises-Fisher models}
\label{sec:GBI}
Based on \cite{bissiri2016general, bissiri2019general}, the general Bayesian framework achieves belief updating while satisfying coherence and additivity for parameters of interest in the $M$-estimation. 
In this framework, belief updating is given through a generalised posterior distribution expressed in a functional form.
This implies that the generalised posterior provides a probabilistic quantification of the uncertainty associated with the $M$-estimator.
Consequently, we apply generalised Bayesian inference to the von Mises-Fisher models using robust divergences.

\subsection{Generalised Posterior distributions based on robust divergences}
In the Bayesian context, the parameter estimation is based on the posterior distribution of parameter $\bm{\xi}$ given by $\bm{x}_{1:n} = (\bm{x}_1, \ldots, \bm{x}_n)$. 
The ordinary posterior density is given as follows: 
\begin{align*}
    \pi(\bm{\xi}\mid\bm{x}_{1:n}) = \frac{L_n(\bm{\xi})\pi(\bm{\xi})}{\int L_n(\bm{\xi})\pi(\bm{\xi})d\bm{\xi}} \propto \exp(-nd_{KL}(\bar{g}, f_{\bm{\xi}}))\pi(\bm{\xi}) ,
\end{align*}
where $L_n(\bm{\xi}) = \prod_{i=1}^{n}f(\bm{x}_i \mid\bm{\xi})$ is the likelihood function and $\pi(\bm{\xi})$ is the prior density of $\bm{\xi}$. 
Since the ordinary posterior corresponds to the specific choice of the negative log-likelihood as a loss function, we adopt divergence-based losses in the generalised Bayesian inference to obtain robust posteriors.
\cite{ghosh2016robust} and \cite{nakagawa2020robust} have proposed generalised posterior distributions $\pi^{(\alpha)}(\bm{\xi}\mid\bm{x}_{1:n})$ and $\pi^{(\gamma)}(\bm{\xi}\mid\bm{x}_{1:n})$ based on DPD and $\gamma$-D, defined as: 
\begin{align}
  \pi^{(\alpha)}(\bm{\xi}\mid\bm{x}_{1:n}) &= \frac{\pi(\bm{\xi})\exp{(-n d_{\alpha}(\bar{g},f_{\bm{\xi}}))}}{\int \pi(\bm{\xi})\exp{(-n d_{\alpha}(\bar{g},f_{\bm{\xi}}))} d\bm{\xi}}, \\
  \pi^{(\gamma)}(\bm{\xi}\mid \bm{x}_{1:n}) &= \frac{\pi(\bm{\xi})\exp{(-n \tilde{d}_{\gamma}(\bar{g},f_{\bm{\xi}}))}}{\int \pi(\bm{\xi})\exp{(-n \tilde{d}_{\gamma}(\bar{g},f_{\bm{\xi}}))} d\bm{\xi}},
\end{align}
where $\tilde{d}_{\gamma}(\bar{g}, f_{\bm{\xi}}) = -\gamma^{-1}\left\{ \exp{(-\gamma d_{\gamma}(\bar{g},f_{\bm{\xi}}))} -1 \right\}$ and $\alpha, \gamma > 0$ is a tuning parameter on robustness.
We refer to these as the DPD posterior and the $\gamma$-D posterior. 
For the von Mises-Fisher distribution, the robust divergence-based $M$-estimators of \cite{kato2016robust} can be naturally extended to the Bayesian context because the empirical cross entropies admit the following additive forms \citep{kato2016robust}:
\begin{align}
d_{\alpha}(\bar{g}, f_{\bm{\xi}}) = n^{-1}\sum_{i = 1}^n \ell_{\alpha}(\bm{x}_i, \bm{\xi}), ~~
 \tilde{d}_{\gamma}(\bar{g}, f_{\bm{\xi}}) = n^{-1}\sum_{i = 1}^n \ell_{\gamma}(\bm{x}_i, \bm{\xi}),    
\end{align}
where 
\begin{align*}
    \ell_{\alpha}(\bm{x}_i, \bm{\xi})& = - \frac{1}{\alpha}\frac{\exp{(\alpha \bm{\xi}^{\top} \bm{x}_i)}}{K_p(\bm{\xi})^\alpha} + \frac{1}{\alpha + 1}\frac{K_p((1 + \alpha)\bm{\xi})}{K_p(\bm{\xi})^{\alpha + 1}}, \\
    \ell_{\gamma}(\bm{x}_i, \bm{\xi}) & = -\frac{1}{\gamma} \frac{\exp{(\gamma \bm{\xi}^{\top} \bm{x}_i)}}{ K_p((1 + \gamma)\bm{\xi})^{{\gamma}/(1+\gamma)} } + \frac{1}{\gamma}. 
\end{align*}
Using this generalised posterior distribution produces a robust estimation for the location and concentration parameters of the von Mises-Fisher distribution, while also providing a probabilistic quantification of uncertainty.

\begin{rem}[Selection of Tuning Parameters]
Selecting the tuning parameters $\alpha$ and $\gamma$ is challenging, and no universally optimal choice is known. Since larger values of $\gamma$ (and similarly $\alpha$) yield stronger robustness, one must balance robustness and efficiency. A common approach to evaluating this trade-off is the asymptotic relative efficiency (ARE). The derivation and detailed discussion of the ARE in this setting are provided in Appendix \ref{app:Select}.  
\end{rem}

\subsection{Asymptotic Properties}
The following asymptotic properties, provided by \cite{ghosh2016robust, nakagawa2020robust}, are also straightforwardly applicable to this situation. 
We now assume the following regularity conditions on the density function $f_{\bm{\xi}}(\bm{x}) = f(\bm{x}; \bm{\xi}) \ (\bm{\xi} \in \mathbb{R}^p)$ and the loss function $\ell_d$. 
Let $\bm{\xi}_g:=\arg \min_{\bm{\xi} \in \mbb{R}^p} \E_g[\ell_d(\bm{X}, \bm{\xi})]$ be the target parameter where $\E_g(\cdot)$ is the expectation of $\bm{X}$ with respect to a probability density function $g$. 
\begin{enumerate}[label = (A\arabic*)]
\item The support of the density function does not depend on the unknown parameter $\bm{\xi}$ and $f_{\bm{\xi}}$ is third order differentiable with respect to $\bm{\xi}$ in the neighbourhood $U$ of $\bm{\xi}_g$. \label{a1}
\item  Interchange of the order of integration with respect to $\bm{x}$ and differentiation as $\bm{\xi}_g$ is justified. The expectations $\E_{g}[\partial_i \ell_d(\bm{X}, {\bm{\xi}}_g)]$ and $\E_{g}[\partial_i\partial_j \ell_d(\bm{X}, {\bm{\xi}}_g)]$ are all finite and $M_{ijk}(\bm{x})$ exists such that \label{a2}
\begin{align*} 
\sup_{\bm{\xi} \in U} \left|\partial_i\partial_j\partial_k \ell_d(\bm{x}, \bm{\xi})\right| \leq M_{ijk}(\bm{x}) \text{ and }  \E_{g}\left[M_{ijk}(\bm{X})\right] < \infty, 
\end{align*}
for all $i, j, k = 1, \ldots, p$, where $\partial_i = \partial/\partial \xi_i$ and $\partial = \partial/\partial \bm{\xi}$.
\item For any $\delta > 0$, with probability one \label{a3}
\begin{align*}
\sup_{\| \bm{\xi} - \bm{\xi}_g \| > \delta } \left\{ d(\bar{g}, f_{\bm{\xi}_g}) - d(\bar{g}, f_{\bm{\xi}}) \right\} < -\ep, 
\end{align*}
for some $\ep > 0$ and for all sufficiently large $n$. 
\end{enumerate}
The matrices $I^{(d)}(\bm{\xi})$ and $J^{(d)}(\bm{\xi})$ are defined by 
 \begin{align*}
I^{(d)}(\bm{\xi}) = \E_{g}\left[\partial \ell_d( \bm{X}; \bm{\xi}) \partial^{\top} \ell_d( \bm{X}; \bm{\xi})\right],~~ J^{(d)}(\bm{\xi}) = \E_{g}\left[\partial \partial^{\top} \ell_d(\bm{X}; \bm{\xi})\right],    
\end{align*}
respectively. We also assume that $I^{(d)}(\bm{\xi})$ and $J^{(d)}(\bm{\xi})$ are positive definite matrices. Under these conditions, \cite{ghosh2016robust} and \cite{nakagawa2020robust} discussed several asymptotic properties of the generalised posterior distributions and the corresponding posterior means.

\begin{thm}[\citealp{ghosh2016robust, nakagawa2020robust}]\label{asymp1}
Under the conditions {\rm \ref{a1}--\ref{a3}}, we assume that $\hat{\bm{\xi}}_n^{(d)}$ is a consistent solution of $\partial d(\bar{g}, f_{\bm{\xi}}) = 0$ and $\hat{\bm{\xi}}^{(d)}_n \xrightarrow{p} \bm{\xi}_g $ as $n \rightarrow \infty$. Then, for any prior density function $\pi(\bm{\xi})$ that is continuous and positive at $\bm{\xi}_g$, it holds that
\begin{align}
\int \left|\pi^{*(d)}(\bm{t_n} \mid \bm{x}_{1:n}) - \phi\left(\bm{t}_n; J^{(d)}(\bm{\xi}_g)^{-1}\right)\right| \dd \bm{t}_n \xrightarrow{p} 0,  \label{siki-main}
\end{align}
as $n \rightarrow \infty$, where $\phi(\cdot; A)$ is the density function of a p-variate normal distribution with zero mean vector and covariance matrix $A$, and $\pi^{*(d)}(\bm{t}_n \mid \bm{x}_{1:n})$ is the generalised posterior density function of the normalized random variable $\bm{t}_n = (t_1, \ldots, t_p)^{\top} = \sqrt{n}(\bm{\xi}-\hat{\bm{\xi}}_n^{(d)})$ given $\bm{x}_{1:n}$. 
\end{thm}

\begin{thm}[\citealp{ghosh2016robust, nakagawa2020robust}]\label{thm2}
In addition to assumptions of Theorem \ref{asymp1}, assume that the prior density $\pi(\bm{\xi})$ has a finite expectation. Let $\tilde{\bm{\xi}}_n^{(d)} = \E^{(d)}(\bm{\xi} \mid \bm{x}_{1:n}) = \int \bm{\xi} \pi^{(d)}(\bm{\xi}\mid \bm{x}_{1:n}) \dd \bm{\xi}$, then it holds that 
\begin{enumerate}[label = (\alph*)]
    \item $\sqrt{n}(\tilde{\bm{\xi}}_n^{(d)} - \hat{\bm{\xi}}_n^{(d)}) \xrightarrow{p} 0$, 
    \item $\sqrt{n}(\tilde{\bm{\xi}}_n^{(d)} - \bm{\xi}_g) \xrightarrow{d} N(\bm{0}, J^{(d)}(\bm{\xi}_g)^{-1}I^{(d)}(\bm{\xi}_g)J^{(d)}(\bm{\xi}_g)^{-1})$. 
\end{enumerate}
\end{thm}
The information matrices $\bm{I}^{(d)}$ and $\bm{J}^{(d)}$ for the ordinary, DPD, and $\gamma$-D posteriors are provided in Section \ref{app:information} of the Appendix.
From Theorem \ref{thm2}, it can be seen that the posterior means of the DPD posterior and the $\gamma$-D posterior are asymptotically equivalent to the estimators that minimize the corresponding empirical cross-entropy. 

\subsection{Posterior Computation}
\label{sec:Comp}

In this section, we consider approximate posterior sampling by optimising randomly weighted likelihood functions using the weighted Bayesian bootstrap (WBB) method.
Generally, sampling from the posterior distribution is typically performed using MCMC methods, such as Hamiltonian Monte Carlo, the Metropolis-Hastings algorithm, and Gibbs sampling. 
In directional statistics, numerous studies have developed MCMC for the von Mises-Fisher distribution \citep{Nunez-antonio2005Bayesian, pal2022modified}, and a general approach to MCMC simulation on embedded Riemannian manifolds was introduced by \cite{byrne2013geodesic, habeck2023geodesic}. 
However, implementing these MCMC methods for the robust divergence-based generalised posterior distributions considered in this study is extremely difficult, because the resulting generalised posterior involves special functions and therefore has a highly complicated form.
Although tools such as Stan \citep{carpenter2017stan} have made MCMC much easier to use in recent years, they cannot be applied in our setting: the DPD and $\gamma$-D posteriors for the von Mises-Fisher models involve Bessel functions of order $(p-2)/2$, and Stan's built-in Bessel functions do not support fractional orders. Therefore, Stan cannot be used when the dimension $p$ is odd.

To address this issue, we consider an approximate sampling strategy for the proposed posterior distributions $\bm{\xi}$ using WBB method proposed by \cite{newton2021weighted}. 
WBB method generates posterior samples as the minimisers of the weighted objective function
\begin{align*}
    L(\bm{\xi}) =  \sum_{i=1}^{n} w_i l(\bm{x}_i;\bm{\xi})  + \lambda w_0 \phi(\bm{\xi}), 
\end{align*}
where $(w_1,\ldots,w_n) \sim n \cdot \text{Dirichlet}(1,\ldots,1)$ are normalised weights satisfying $\sum_{i=1}^n w_i=n$, and $w_0$ is distributed as a univariate Exponential.  
The penalty function (or regularisation term) $\lambda w_0 \phi(\bm{\xi})$ imposes a constraint on the parameter $\bm{\xi}$, and its strength is controlled by the hyperparameter $\lambda > 0$.  
A typical example of such a penalty function is the negative log prior with $\lambda = 1$, which allows incorporating prior information into the objective function.
The sampling algorithm is presented below.
\begin{algorithm}[H]
\caption{Weighted Bayesian Bootstrap (WBB) Algorithm}
\label{alg:WBB}

Repeating the following steps 1--4 for $m = 1,\ldots,M$ yields the posterior samples 
$\bigl(\widehat{\bm{\xi}}^{(1)},\ldots,\widehat{\bm{\xi}}^{(M)}\bigr)$.
\begin{enumerate}
    \item Generate $e_i^{(m)} \sim \mathrm{Exp}(1)$ for $i=1,\ldots,n$, and normalize them to obtain the weights $w_i^{(m)} = n e_i^{(m)}/\sum_{j = 1}^ne_j^{(m)}$.
    \item Generate the penalty weight $w_0^{(m)} \sim \mathrm{Exp}(1)$.
    \item Construct the weighted objective function $L^{(m)}(\bm{\xi})$.
    \item Compute the optimizer $\widehat{\bm{\xi}}^{(m)} = \arg\min_{\bm{\xi}} L^{(m)}(\bm{\xi})$.
\end{enumerate}
\end{algorithm}

WBB method is an optimisation-based approach, and in this paper we employ the \texttt{optim} function in R.

\begin{figure}[htbp]
    \centering
        \includegraphics[width=0.90\textwidth]{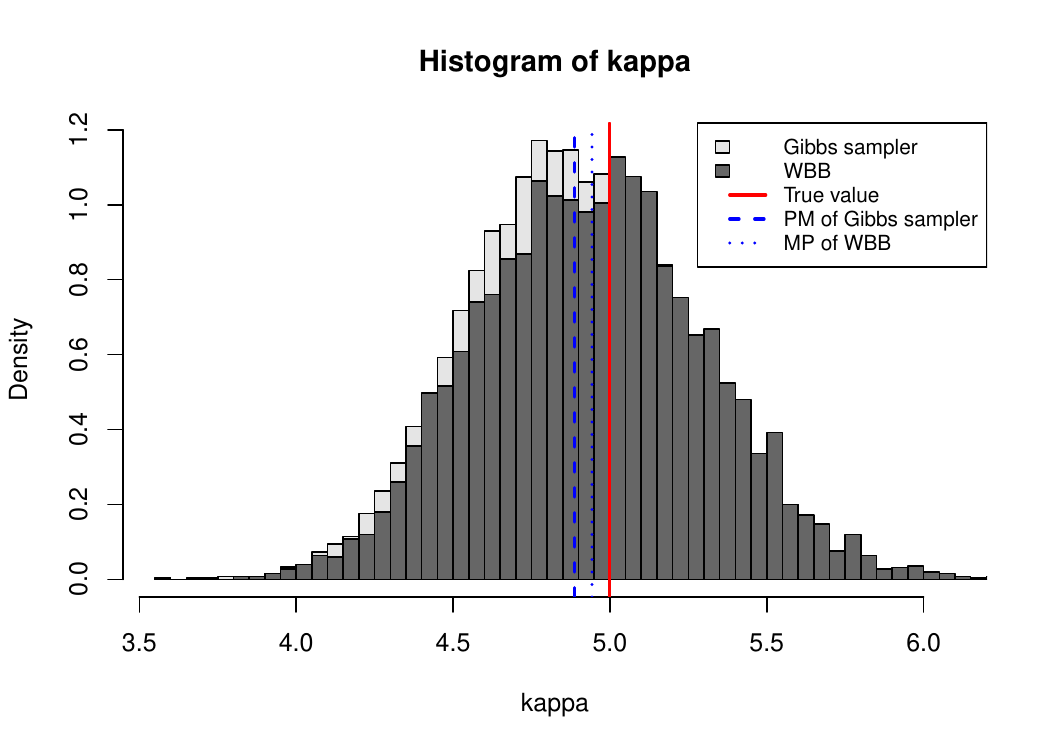} 
    \caption{Histograms of WBB and Gibbs sampler in the case of $p = 3$. }
    \label{fig:gibbs}
\end{figure}

In Figure \ref{fig:gibbs}, we compare the performance of WBB and the Gibbs sampler proposed by \cite{pal2022modified} in terms of sampling from the ordinary posterior distribution in the absence of outliers. 
From Figure \ref{fig:gibbs}, it can be seen that the results obtained by WBB are comparable to those of the Gibbs sampler.

\section{Robustness}
\label{sec:Robust}

Similarly to \cite{ghosh2016robust} and \cite{nakagawa2020robust}, we consider the robustness of the posterior mean based on the DPD posterior and the $\gamma$-D posterior in terms of the Bayesian influence function proposed in \cite{hooker2014bayesian}. 
The Influence Function (IF) is a measure used in robust statistics to evaluate the robustness of an estimator in terms of its sensitivity to small changes in the data.
However, IF is bounded because the directional data is on a compact set. In such a case, we consider the Standardised Influence Function (SIF). 
The SIF provides a normalised version of the IF to ensure scale and parameterisation invariance, making it particularly suitable for comparing robustness across different models and scenarios \citep[for example, ][so on.]{ko1988robustness, ko1993robust, kato2016robust}.
Let $G$ denote the data-generating distribution with density $g$ and consider the parametric family $\{F_{\bm{\xi}}:\bm{\xi} \in \mbb{R}^{p}\}$ with density $f_{\bm{\xi}}$. Let $\pi(\bm{\xi})$ be the prior density for $\bm{\xi}$. The DPD posterior and $\gamma$-D posterior densities as a function of $G$ and $\bm{\xi}$ are defined by
\begin{align}
\pi^{(\alpha)}(\bm{\xi};G) & =\frac{\exp(nQ^{(\alpha)}(\bm{\xi}; G,F_{\bm{\xi}}) )\pi(\bm{\xi})}{\int \exp(nQ^{(\alpha)}(\bm{\xi}; G,F_{\bm{\xi}}) )\pi(\bm{\xi}) \dd\bm{\xi}}, \\
\pi^{(\gamma)}(\bm{\xi};G)& =\frac{\exp(nQ^{(\gamma)}(\bm{\xi}; G,F_{\bm{\xi}}) )\pi(\bm{\xi})}{\int \exp(nQ^{(\gamma)}(\bm{\xi}; G,F_{\bm{\xi}}) )\pi(\bm{\xi}) \dd\bm{\xi}},
\end{align}
where
\begin{align}
 Q^{(\alpha)}(\bm{\xi}; G,F_{\bm{\xi}}) &= \frac{1}{\alpha}\int  f_{\bm{\xi}}(\bm{x})^{\alpha} dG(\bm{x}) -\frac{1}{1+\alpha} \int f_{\bm{\xi}}(\bm{x})^{1+\alpha}\dd x, \\ 
 Q^{(\gamma)}(\bm{\xi}; G,F_{\bm{\xi}}) &= \frac{1}{\gamma}\left[\int  f_{\bm{\xi}}(\bm{x})^{\gamma} dG(\bm{x}) \right]\left(\int f_{\bm{\xi}}(\bm{x})^{1+\gamma} \dd \bm{x} \right)^{-\gamma/ (1+\gamma)}.
 \end{align}
For a fixed sample size $n$, the DPD Bayes functional and $\gamma$-D Bayes functional under the general loss function $L(\cdot, \cdot)$ are given as
\begin{align}
T_n^{(\alpha)L}(G)=\arg \min_{\bm{t}} \frac{\int L(\bm{\xi},\bm{t})\exp(nQ^{(\alpha)}(\bm{\xi}; G,F_{\bm{\xi}}) )\pi(\bm{\xi})\dd\bm{\xi}}{\int \exp(nQ^{(\alpha)}(\bm{\xi}; G,F_{\bm{\xi}}) )\pi(\bm{\xi}) \dd\bm{\xi}}, \\
T_n^{(\gamma)L}(G)=\arg \min_{\bm{t}} \frac{\int L(\bm{\xi},\bm{t})\exp(nQ^{(\gamma)}(\bm{\xi}; G,F_{\bm{\xi}}) )\pi(\bm{\xi})\dd\bm{\xi}}{\int \exp(nQ^{(\gamma)}(\bm{\xi}; G,F_{\bm{\xi}}) )\pi(\bm{\xi}) \dd\bm{\xi}}.
\end{align}
Under the quadratic loss function, the DPD Bayes functional $T_n^{(\alpha)L}(G)$ and the $\gamma$-D Bayes functional $T_n^{(\gamma)L}(G)$ are the DPD posterior mean functional $T_{n}^{(\alpha)e}(G)$ and the $\gamma$-D posterior mean functional $T_{n}^{(\gamma)e}(G)$, respectively. 
$T_{n}^{(\alpha)e}(G)$ and $T_{n}^{(\gamma)e}(G)$ are given as 
\begin{align*}
T_{n}^{(\alpha)e}(G)&=\frac{\int \bm{\xi}\exp(nQ^{(\alpha)}(\bm{\xi}; G,F_{\bm{\xi}}) )\pi(\bm{\xi})d\bm{\xi}}{\int \exp(nQ^{(\alpha)}(\bm{\xi}; G,F_{\bm{\xi}}) )\pi(\bm{\xi}) d\bm{\xi}}, \\
T_{n}^{(\gamma)e}(G)&=\frac{\int \bm{\xi}\exp(nQ^{(\gamma)}(\bm{\xi}; G,F_{\bm{\xi}}) )\pi(\bm{\xi})d\bm{\xi}}{\int \exp(nQ^{(\gamma)}(\bm{\xi}; G,F_{\bm{\xi}}) )\pi(\bm{\xi}) d\bm{\xi}}. 
\end{align*}
Hereafter, we consider the case of the quadratic loss function. We now consider the contaminated model $F_{\varepsilon}=(1-\varepsilon)G+\varepsilon \Delta_{\bm{y}}$, where $\varepsilon$ is the contamination ratio and $\Delta_{\bm{y}}$ is the degenerate contaminating distribution at $\bm{y}$. Then the influence functions of the DPD posterior mean functional and the $\gamma$-D posterior mean functional for a fixed $n$ at $G$ are defined by
\begin{align}\label{IFgamma}
\textrm{IF}_n (\bm{y},T_n^{(\alpha)e},G)&=\frac{\partial}{\partial \varepsilon} T_n^{(\alpha)e}(F_{\varepsilon}) \bigg|_{\varepsilon=0}=n {\rm Cov}_{\pi^{(\alpha)}(\bm{\xi};G)}(\bm{\xi}, k_{\alpha}(\bm{\xi};\bm{y},g)),\\
\textrm{IF}_n (\bm{y},T_n^{(\gamma)e},G)&=\frac{\partial}{\partial \varepsilon} T_n^{(\gamma)e}(F_{\varepsilon}) \bigg|_{\varepsilon=0}=n {\rm Cov}_{\pi^{(\gamma)}(\bm{\xi};G)}(\bm{\xi}, k_p{\gamma}(\bm{\xi};\bm{y},g)),
\end{align}
where ${\rm Cov}_{\pi(\bm{\xi};G)}$ is the covariance under the posterior $\pi(\bm{\xi};G)$ and 
\begin{align}\label{Qgamma}
k_{\alpha}(\bm{\xi};\bm{y},g) &=\frac{\partial}{\partial \varepsilon} Q^{(\alpha)}(\bm{\xi};F_{\varepsilon},F_{\bm{\xi}})=\frac{1}{\alpha}\left[ f_{\bm{\xi}}(\bm{y})^{\alpha}-\int f_{\bm{\xi}}(\bm{x})^{\alpha}g(\bm{x})\dd \bm{x} \right], \\
k_{\gamma}(\bm{\xi};\bm{y},g) &=\frac{\partial}{\partial \varepsilon} Q^{(\gamma)}(\bm{\xi};F_{\varepsilon},F_{\bm{\xi}})=\frac{1}{\gamma}\left[ f_{\bm{\xi}}(\bm{y})^{\gamma}-\int f_{\bm{\xi}}(\bm{x})^{\gamma}g(\bm{x})\dd \bm{x} \right] C_{\bm{\xi}, \gamma},  
\end{align}
for $\alpha, \gamma>0$, where $C_{\bm{\xi}, \gamma} = \left( \int f_{\bm{\xi}}(\bm{x})^{1+\gamma } \dd \bm{x}\right)^{-\gamma/(1+\gamma)}$. For $\alpha = \gamma=0$, we have $k_0(\bm{\xi};\bm{y},g)=\log f_{\bm{\xi}}(\bm{y})-\int g(\bm{x}) \log f_{\bm{\xi}}(\bm{x}) \dd\bm{x}$ which is the influence function of the ordinary posterior mean for a fixed $n$. 

When $g(x) = f(\bm{x}\mid \bm{\eta})$, $k_{\alpha}(\bm{\xi};\bm{y},g)$ and $k_{\gamma}(\bm{\xi};\bm{y},g)$ are given by 
\begin{align}
    k_{\alpha}(\bm{\xi};\bm{y},g) &= \frac{1}{\alpha}\frac{1}{K_p(\bm{\xi})^{\alpha}}\left[ \exp{(\alpha\bm{\xi}^{\top}\bm{y})} - \frac{K_p(\alpha \bm{\xi} + \bm{\eta})}{K_p(\|\bm{\eta}\|)}\right], \\    
    k_{\gamma}(\bm{\xi};\bm{y},g) &= \frac{1}{\gamma}\frac{1}{K_p((1 + \gamma)\bm{\xi})^{\gamma/(1 + \gamma)}}\left[ \exp{(\gamma\bm{\xi}^{\top}\bm{y})} - \frac{K_p(\gamma \bm{\xi} + \bm{\eta})}{K_p(\|\bm{\eta}\|)}\right]. 
\end{align}
For $\alpha = \gamma=0$, we have $k_0(\bm{\xi}; \bm{y},g)= \bm{\xi}^{\top}\left\{\bm{y} -(A_p(\|\bm{\eta}\|)/\|\bm{\eta}\| )\bm{\eta}\right\}$. 

 Then the norm of SIF is given by 
\begin{align*}
\textrm{SIF}(\bm{y},T_n, G) = \sqrt{\textrm{IF}(\bm{y},T_n, G)^{\top}\bm{S}(G)^{-1}\textrm{IF}(\bm{y},T_n, G)}. 
\end{align*}

\begin{figure}[htbp]
    \centering
    \subfigure[Ordinary posterior]{
        \includegraphics[width=0.30\textwidth]{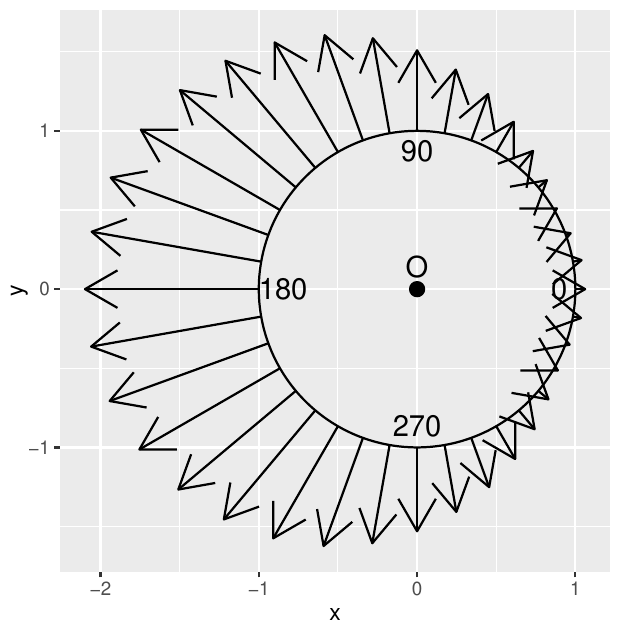} 
    }
    \subfigure[DPD posterior ($\alpha = 0.15$)]{
        \includegraphics[width=0.30\textwidth]{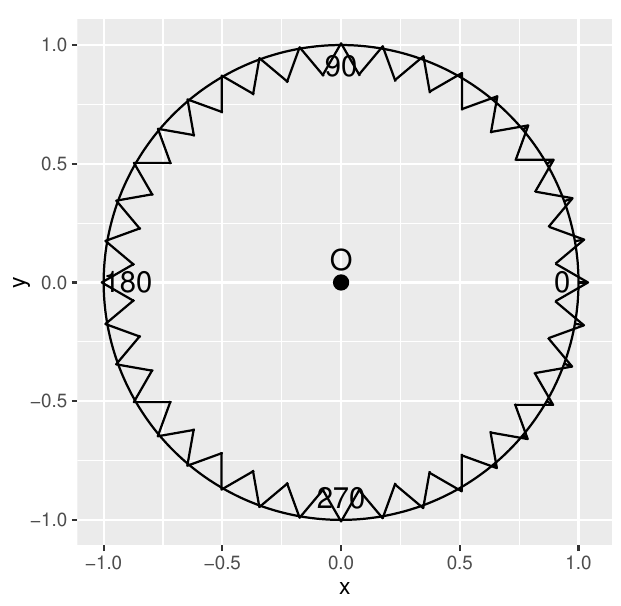} 
    }
    \subfigure[$\gamma$-D posterior ($\gamma = 0.15$)]{
        \includegraphics[width=0.30\textwidth]{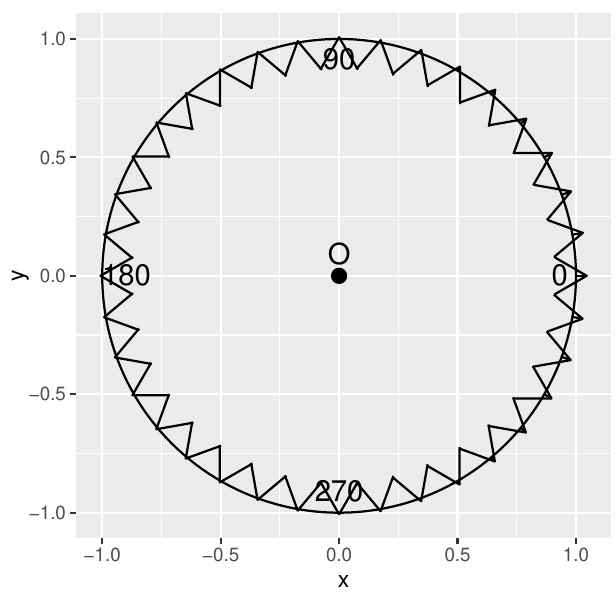} 
    }
   \caption{The figures display the SIF for these posterior means.  The arrow length is value of the norm of SIF, and the direction of arrow indicates the direction of $\bm{y}$. The setting of this figures are that distribution $G$ is von Mises-Fisher distribution $vM(\bm{\mu}_0, \kappa_0)$ with $\bm{\mu}_0 = (1, 0)^{\top}$, $\kappa_0 = 5$, $\alpha = 0.15$, and $\gamma = 0.15$. }
    \label{fig:influence}
\end{figure}

Figure \ref{fig:influence} illustrates $\textrm{SIF}(\bm{y},T_n, G)$. In these graphs, the length of the arrow is a value of $\textrm{SIF}(\bm{y},T_n, G)$ and the direction of the arrow indicates the direction of $\bm{y}$. 
It shows that the ordinary posterior mean is highly sensitive to the presence of an opposite-direction observation for $\bm{\mu}_0 = (1, 0)^{\top}$. 
By contrast, the DPD posterior and $\gamma$-D posterior means remain stable and are not influenced by such observations. 

\section{Simulation Studies}
\label{sec:sim}

In this section, we evaluate and compare the performance of estimators in numerical studies based on the mean squared error (MSE). 
We generate random samples of sizes $n = 100, 150, 200$, dimensions $p = 2, 3, 5$, and contamination ratios $\ep = 0.00, 0.05, 0.10$ from the contaminated distribution $(1 - \ep)f_{\bm{\xi}}(\bm{x}) + \ep u(\bm{x})$ where $f_{\bm{\xi}}(\bm{x})$ is the probability density function of the von Mises-Fisher distributions $\text{vM}_p(\bm{\xi})$ and $u(\bm{x})$ is the contamination distribution. 
We conduct $300$ replications for each combination of sample size, dimension, and the contamination ratio under uniform prior distribution.  
The performance of the estimators is assessed in terms of MSE, defined as $\sum_{j=1}^{300}\|\hat{\bm{\xi}}_j-\bm{\xi}\|^2/300p$, $\sum_{j=1}^{300} (1 - \hat{\bm{\mu}}_j^{\top}\bm{\mu})/300$, and $\sum_{j=1}^{300}(\hat{\kappa}_j-\kappa)^2/300$ where $\hat{\bm{\xi}}_j$, $\hat{\bm{\mu}}_j$, and $\hat{\kappa}_j$ ($j=1,\dots,300$) are the estimates of $\bm{\xi}$, $\bm{\mu}$, and $\kappa$ for the $j$th simulation sample, respectively.

\begin{figure}[htbp]
    \centering
     \includegraphics[width=1.0\textwidth]{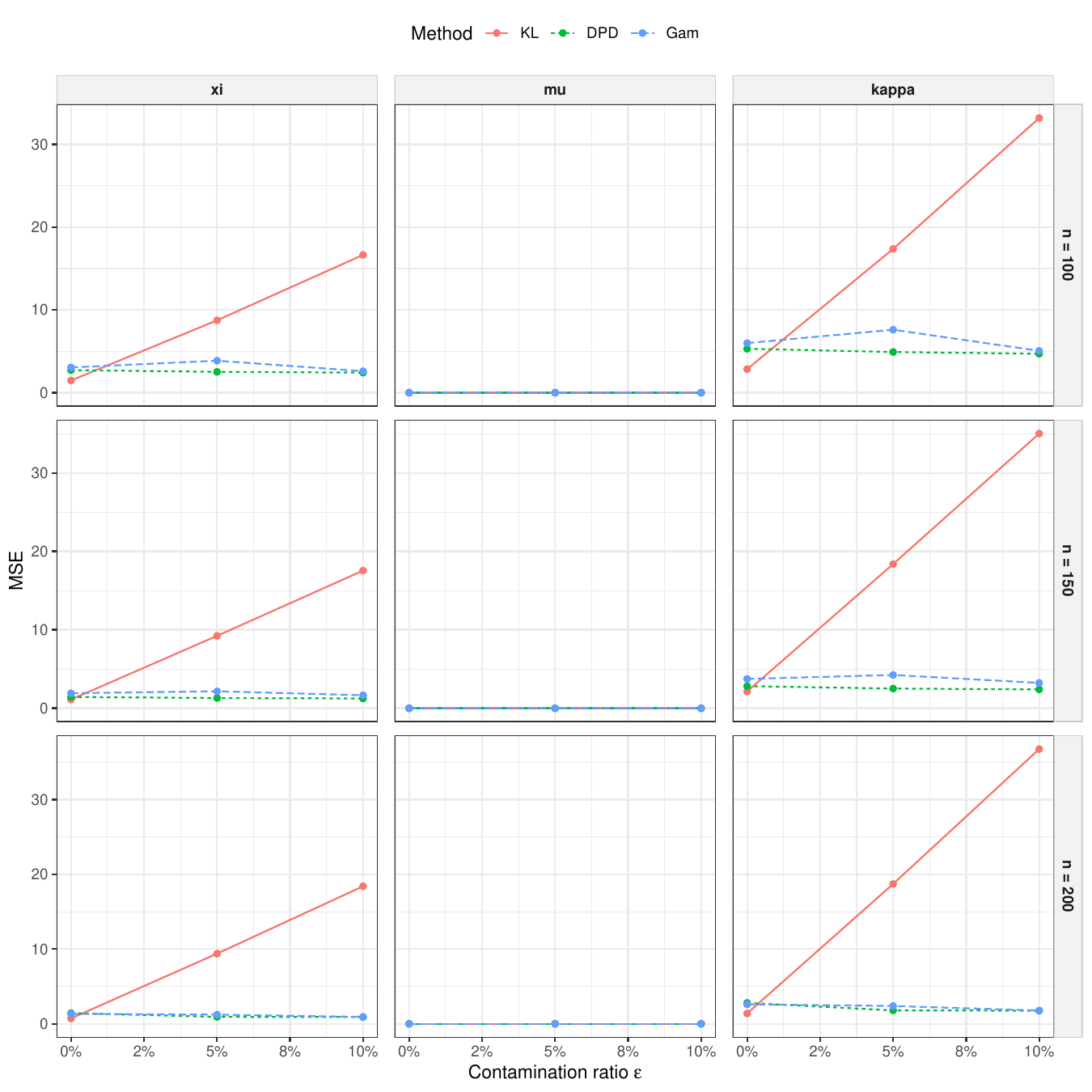}
    
   \caption{These figures show the mean squared errors (MSEs) for the parameters of the von Mises-Fisher distribution with $p = 2$. Rows correspond to the sample sizes $n$, while columns list the parameters $\bm{\xi}$, $\bm{\mu}$, $\kappa$.  The red solid line (KL), the green dotted line (DPD) and the blue dash line (Gam) represent the MSEs of the ordinary posterior, DPD posterior, and $\gamma$-D posterior means using a uniform prior for $\bm{\xi}$, respectively. Outliers are generated from the uniform distribution on the unit sphere.}
    \label{uni_uni:p2}
\end{figure}

\begin{figure}[htbp]
    \centering
     \includegraphics[width=1.0\textwidth]{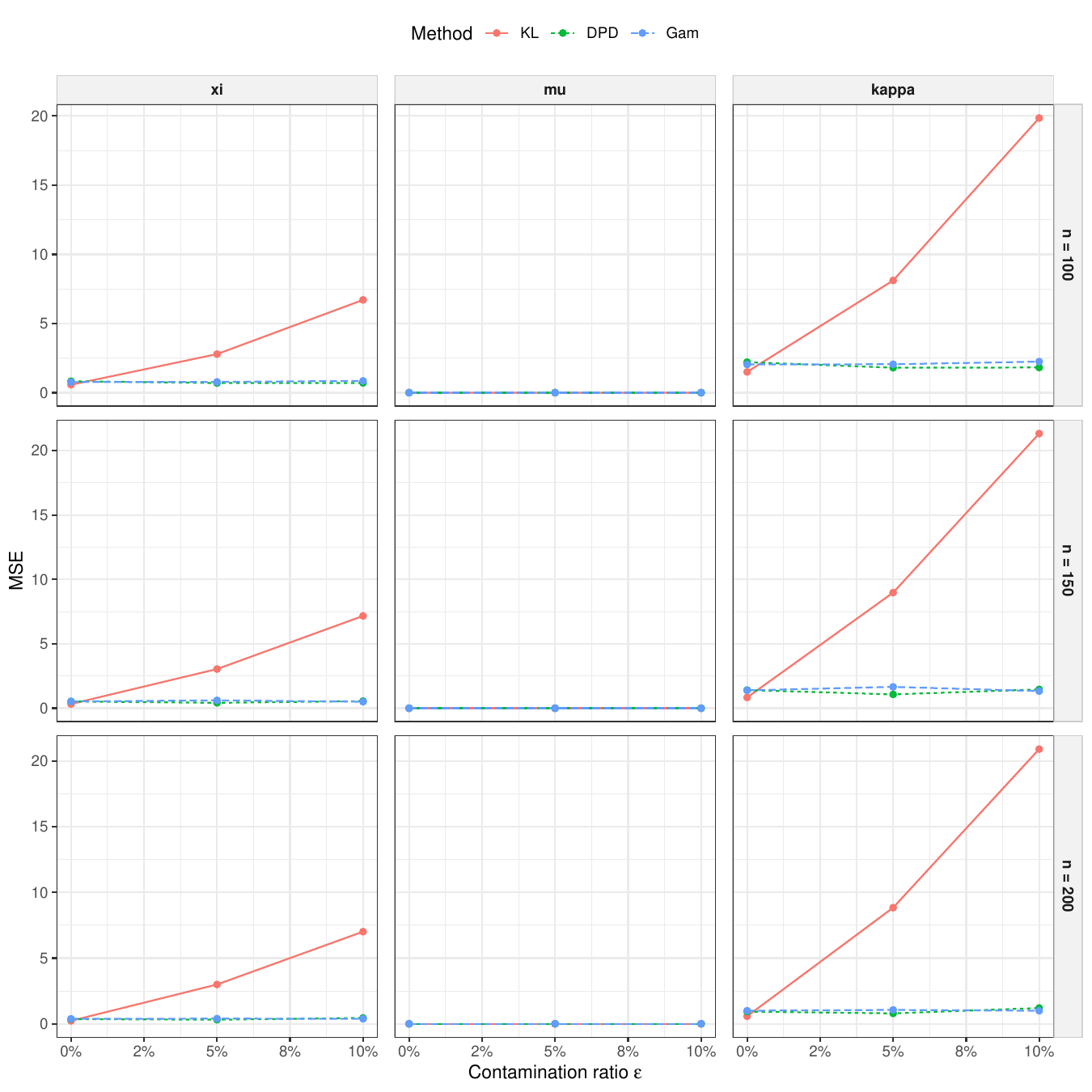}
    
   \caption{These figures show the mean squared errors (MSEs) for the parameters of the von Mises-Fisher distribution with $p = 3$. Rows correspond to the sample sizes $n$, while columns list the parameters $\bm{\xi}$, $\bm{\mu}$, $\kappa$.  The red solid line (KL), the green dotted line (DPD) and the blue dash line (Gam) represent the MSEs of the ordinary posterior, DPD posterior, and $\gamma$-D posterior means using a uniform prior for $\bm{\xi}$, respectively. Outliers are generated from the uniform distribution on the unit sphere.}
    \label{uni_uni:p3}
\end{figure}

\begin{figure}[htbp]
    \centering
     \includegraphics[width=1.0\textwidth]{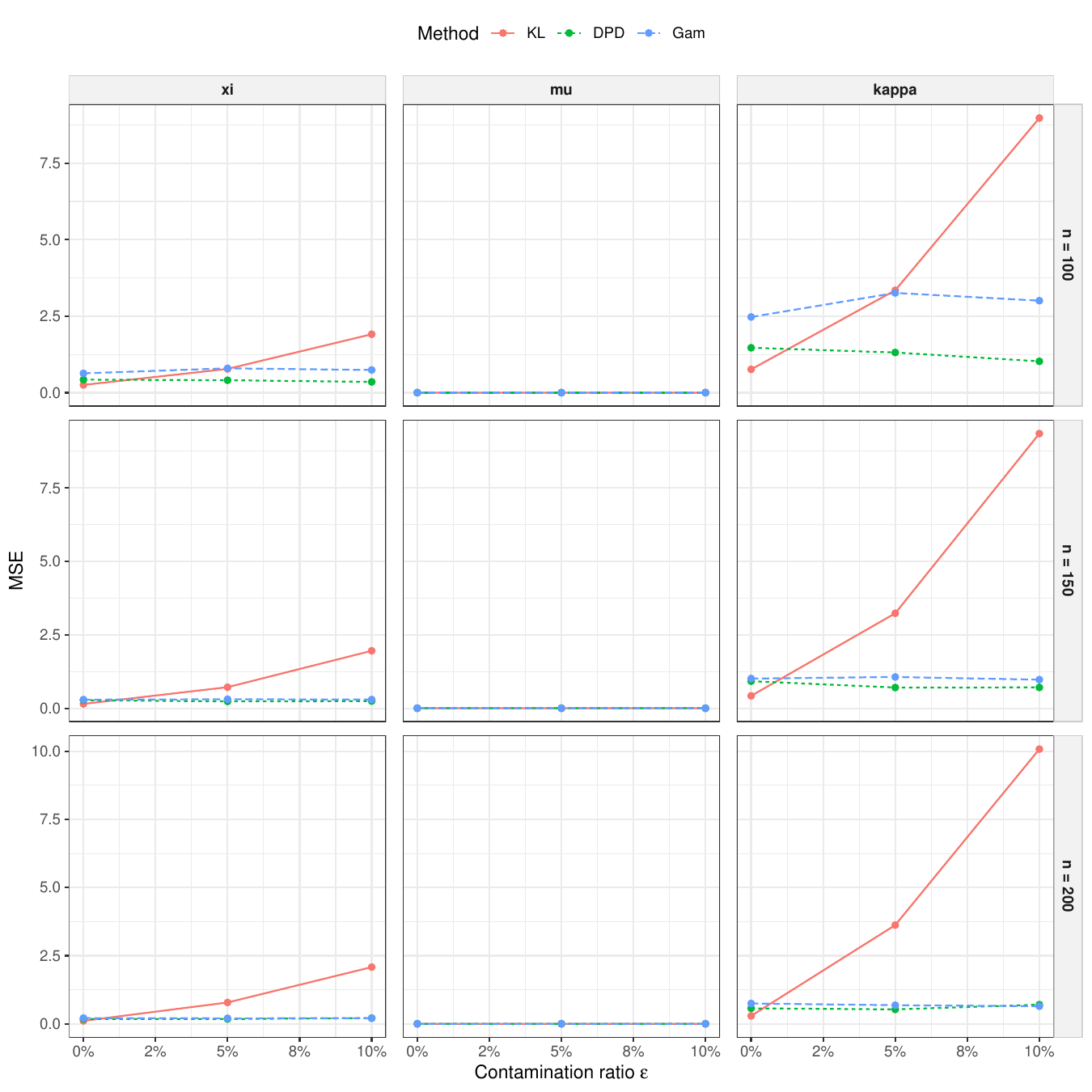}
    
   \caption{These figures show the mean squared errors (MSEs) for the parameters of the von Mises-Fisher distribution with $p = 5$. Rows correspond to the sample sizes $n$, while columns list the parameters $\bm{\xi}$, $\bm{\mu}$, $\kappa$.  The red solid line (KL), the green dotted line (DPD) and the blue dash line (Gam) represent the MSEs of the ordinary posterior, DPD posterior, and $\gamma$-D posterior means using a uniform prior for $\bm{\xi}$, respectively. Outliers are generated from the uniform distribution on the unit sphere.}
    \label{uni_uni:p5}
\end{figure}

Figures \ref{uni_uni:p2}--\ref{uni_uni:p5} present experiments evaluating MSE under a uniform prior for the parameter $\bm{\xi}$ in dimensions $p = 2, 3,$ and $5$, respectively, where the contamination distribution $u(\bm{x})$ is the uniform distribution on $\mcal{S}_{p}$. 
In these figures, the red solid line (KL), the green dotted line (DPD) and the blue dashed line (Gam) represent MSEs of the ordinary posterior mean, DPD posterior mean, and the $\gamma$-D posterior mean, respectively. 
In addition, the tuning parameters were selected using ARE criterion described in Appendix\ref{app:Select}, choosing the values that retain about 95\% of the efficiency of the ordinary posterior mean.

From Figures \ref{uni_uni:p2}--\ref{uni_uni:p5}, we can see that MSE of the ordinary posterior mean (KL) increases as the contamination ratio increases.
In contrast, the DPD posterior mean and the $\gamma$-D posterior mean remain unaffected across all outlier contamination ratios.
In the absence of outliers, MSEs of all posterior means decrease with increasing sample size confirm that they are consistent estimators. 
On the other hand, in the presence of outliers, the ordinary posterior mean fails to be consistent, since its MSE does not decrease with larger sample sizes. 
In contrast, MSEs of the DPD posterior mean and the $\gamma$-D posterior means remain decreasing despite the presence of outliers. 
From Figure \ref{uni_uni:p5}, we can see that, for small samples and high dimensions, the MSE of the $\gamma$-D posterior mean for $\kappa$ is larger than that of DPD posterior mean. 
Thus, the DPD posterior mean shows better performance under such contaminated settings, although the difference in MSE between the DPD and $\gamma$-D posteriors diminishes as the sample size increases.
Therefore, we show that the ordinary posterior mean (KL) is more sensitive to the outliers than the DPD posterior (DPD) and $\gamma$-D posterior (Gam) means.

\section{Application to real data}
\label{sec:DA}
In this section, we apply the proposed methods to two real datasets
to illustrate their behaviour in real data applications.
The R code supporting the results presented in this section is available at
\url{https://github.com/nakagawa96/robust-gb-vmf}.
\subsection{Wind direction data}
\begin{figure}[htbp]
    \centering
     \includegraphics[width=0.8\textwidth]{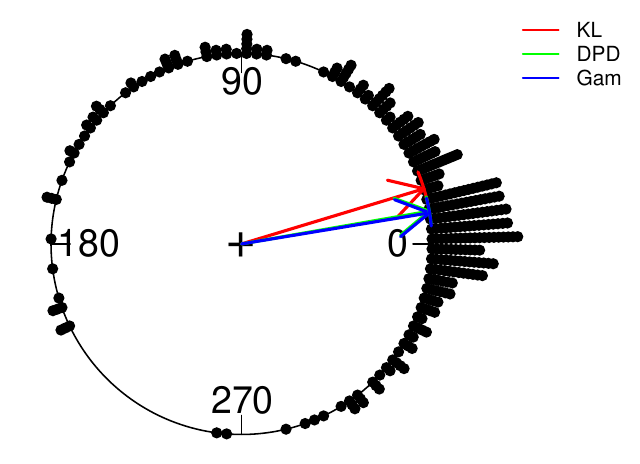}
    
   \caption{Each arrow represents the estimated posterior mean direction on the unit circle, while the surrounding points indicate the sample observations.
The red (KL, $0.29$ rad), green (DPD, $0.17$ rad), and blue (Gamma, $0.16$ rad) arrows show the posterior mean directions, and the arcs denote the 95\% credible intervals of $\theta = \arctan(\mu_2/\mu_1)$, which are $(0.20, 0.39)$, $(0.10, 0.25)$, and $(0.09, 0.24)$, respectively.}
    \label{realdata1}
\end{figure}

We used the \texttt{wind} dataset provided in the \texttt{circular} R package \citep{R-circular}
to illustrate the basic characteristics of directional data.
From Figure \ref{realdata1}, while most observations are concentrated around $0$ radians,
a non-negligible number of observations lie between $\pi/2$ and $\pi$ radians and may be
regarded as potential outlying observations relative to the main concentration.
Figure \ref{realdata1} illustrates the posterior mean directions of $\bm{\mu} = (\mu_1, \mu_2)^{\top}$ estimated from real data using three different posteriors: the ordinary posterior (KL), the DPD posterior, and the $\gamma$-D posterior. 
Each arrow represents the estimated posterior mean direction on the unit circle, while the surrounding points indicate the sample observations. 
The tuning parameters $\alpha$ and $\gamma$ are chosen so that ARE is $95\%$, see Appendix \ref{app:Select} for details.

From Figure \ref{realdata1}, we can see that the ordinary posterior mean (KL; $0.29$ radian) is noticeably shifted due to the influence of outliers, whereas the DPD posterior mean (DPD; $0.17$ radian) and $\gamma$-D posterior mean (Gam; $0.16$ radian) remain stable and close to the dominant direction of the data, demonstrating their robustness.
The circular dispersion of the posterior samples for $\bm{\mu}$ was smallest for the DPD and $\gamma$-D posteriors, with mean resultant lengths of $0.9993$ and $0.9993$, compared to $0.9989$ for the KL posterior.
Consistently, the estimated concentration parameters $\kappa$ were higher for the DPD and $\gamma$-D posteriors ($2.99$ and $3.69$, respectively) than for the ordinary posterior ($1.78$), confirming their robustness to potential outlying observations.

\subsection{Gene expression data}
Directional data also arise naturally in bioinformatics \citep[e.g.][]{eisen1998cluster}, particularly in the analysis of gene expression data, in which the von Mises model has been applied for modelling directional structures \citep[e.g.][]{banerjee2005clustering}. 
We used yeast gene expression data from the NCBI Gene Expression Omnibus (accession GSE12345) \citep{edgar2002gene}, retrieved via the R package GEOquery \citep{davis2007GEOquery}, and processed using Biobase from the Bioconductor project \citep{Gentleman2004Bioconductor}. 
To interpret gene expression profiles as directional data, each observation $\bm{z}_i = (z_{i1}, \ldots, z_{ip})^{\top}$ is normalised to a vector $\bm{x}_i = (x_{i1}, \ldots, x_{ip})^{\top}$ on the unit sphere by
\[
x_{ik} 
= \frac{z_{ik} - \bar{z}_i}
{\sqrt{\sum_{k=1}^{p} (z_{ik} - \bar{z}_i)^2}},~~(i = 1, \ldots, n), 
\]
where $\bar{z} = p^{-1}\sum_{k=1}^{p} z_{ik}$.
This normalisation ensures that $|\bm{x}_i|_2 = 1$, so that each sample lies on the unit sphere.
Under this transformation, the Pearson correlation between $\bm{z}_i$ and $\bm{z}_j)$ can be equivalent to the cosine similarity between $\bm{x}_i$ and $\bm{x}_j$. 

\begin{figure}[htbp]
    \centering
     \includegraphics[width=0.8\textwidth]{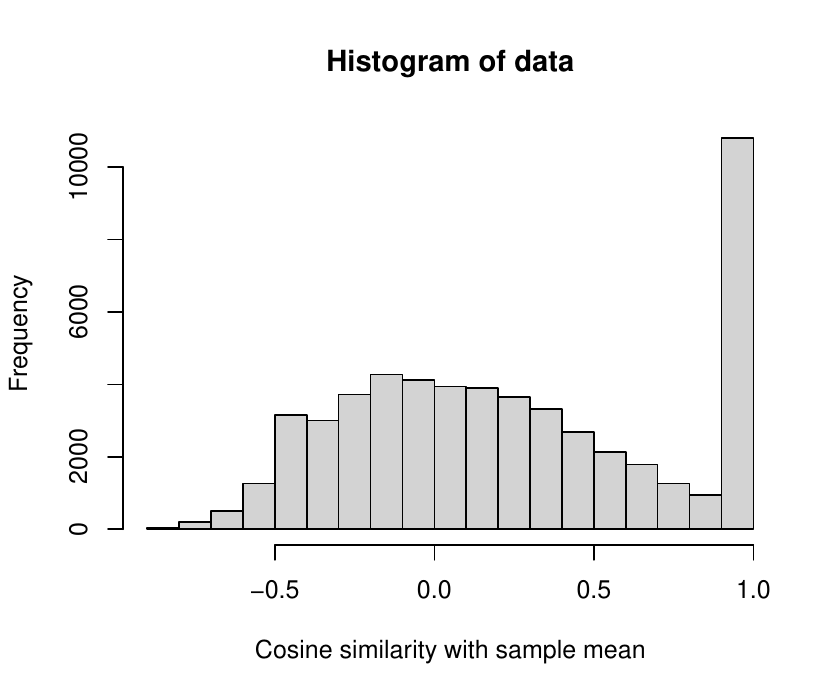}
    
   \caption{Figure displays the histogram of the cosine similarities between data and sample mean. }
    \label{realdata2}
\end{figure}

Figure \ref{realdata2} presents the histogram of the cosine similarities between the data and the sample mean direction $\bar{\bm{x}}/\|\bar{\bm{x}}\|$ where $\bar{\bm{x}} = n^{-1}\sum_{i = 1}^n \bm{x}_i$. 
This figure shows that while most observations lie near the sample mean direction, a non-negligible number of points are widely scattered. 

\begin{figure}
\begin{minipage}[t]{.32\textwidth}
\centering
\includegraphics[width=4.0cm]{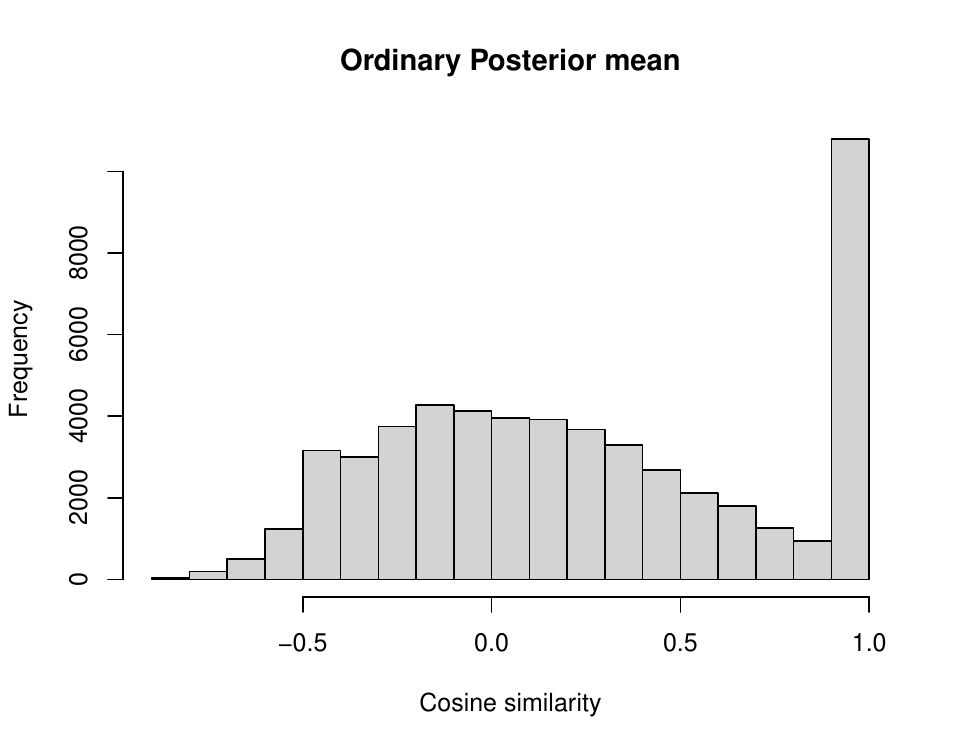}
\end{minipage}
\begin{minipage}[t]{.32\textwidth}
\centering
\includegraphics[width= 4.0cm]{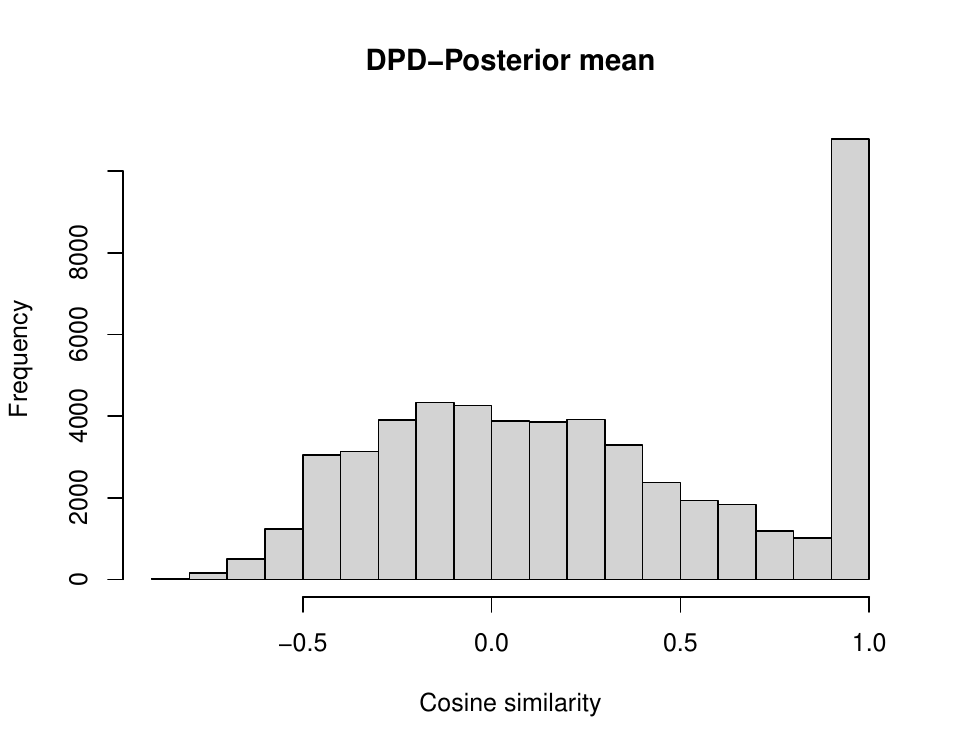}
\end{minipage}
\begin{minipage}[t]{.32\textwidth}
\centering
\includegraphics[width= 4.0cm]{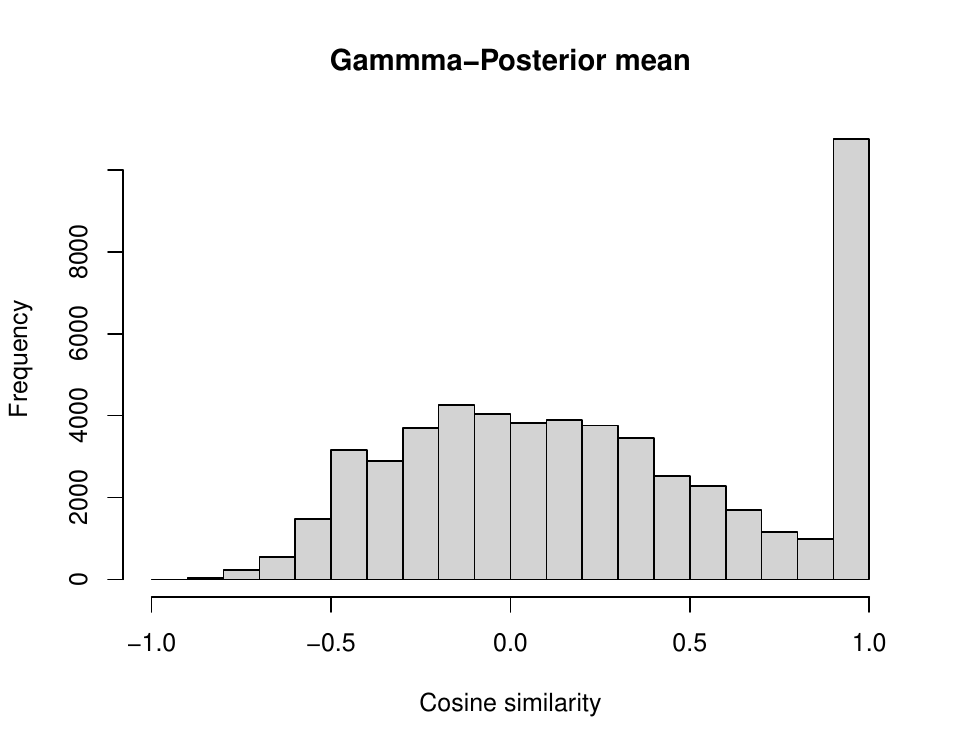}
\end{minipage}
\caption{The figures display histograms of the cosine similarities between the data and the posterior mean directions for the ordinary, DPD, and $\gamma$-D posteriors, respectively.}
\label{fig_realdata2}
\end{figure}

Table \ref{table_realdata2_1} presents the angular distance and cosine distance between posterior mean directions obtain from the ordinary, DPD, and $\gamma$-D posteriors ($\alpha = \gamma =  0.15$). 
Figure \ref{fig_realdata2} also provides histograms of the cosine similarities between the data and these posterior mean directions. 
From the table and figure, we can see that the posterior mean directions are close to one another, suggesting minimal differences in the resulting point estimates. 
On the other hand, Table \ref{table_realdata2_2}, which summarises the posterior means of $\kappa$, shows that the ordinary posterior markedly underestimates $\kappa$ due to the strong influence of outliers.
By contrast, the DPD and $\gamma$-D posteriors yield considerably more robust estimates.

\begin{table}[t]
\centering
\caption{Angular distances and cosine distances between posterior mean directions obtained from the ordinary, DPD, and $\gamma$-D posteriors.}
\label{table_realdata2_1}
\begin{tabular}{ccc}
\toprule
Pair & Angle distance (radian) & cosine distance  \\
\midrule
Ordinary vs DPD & 0.1822988 & 0.01657046 \\
Ordinary vs $\gamma$-D & 0.1721023 & 0.01477308 \\
DPD vs $\gamma$-D & 0.1741671 & 0.01512879 \\
\bottomrule
\end{tabular}
\end{table}

\begin{table}[t]
\caption{Comparison of posterior estimates of the concentration parameter $\kappa$ for the Ordinary, DPD, and $\gamma$-D posteriors.}
\label{table_realdata2_2}
\centering
\begin{tabular}{lccc}
\toprule
 & Ordinary & DPD & $\gamma$-D \\
\midrule
$\kappa$ & 3.3450 & 177.0471 & 159.6273 \\
\bottomrule
\end{tabular}
\end{table}


\section{Conclusion}
\label{sec:summary}

In this paper, we have developed a robust generalised Bayesian inference for von Mises-Fisher models based on the density power divergence and the $\gamma$-divergence. 
The proposed generalised posterior distributions provide a principled way to incorporate robustness into Bayesian estimation for both the location and concentration parameters, where the presence of outliers can severely distort ordinary Bayesian inference.

We established the asymptotic properties of the resulting posterior means, showing that they are asymptotically equivalent to the corresponding divergence-based $M$-estimators. The standardized influence function analysis also demonstrated that the DPD and $\gamma$-D posterior means are robust, remaining stable even in the presence of observations oriented in the opposite direction to the main direction. By contrast, the ordinary posterior mean is highly sensitive to such opposite-direction outliers.

To address the computational challenges arising from Bessel functions and the lack of suitable MCMC methods for fractional-order Bessel terms, we adopted WBB for posterior computation. Numerical comparisons with a Gibbs sampler for the ordinary posterior confirmed that WBB provides a computationally efficient and accurate approximation, even in moderately high-dimensional settings.
Although the density power divergence involves an integral term in its loss function, several strategies for handling cases in which this integral does not admit a closed-form expression have been discussed in recent studies \citep[e.g.,][]{sonobe2025sampling}.
Consequently, these developments also enable straightforward extensions to more complex models.

Simulation studies across multiple dimensions, sample sizes, and contamination ratios revealed the clear advantage of the proposed methods. While the MSE of the ordinary posterior mean increases sharply as the contamination ratio increases, both the DPD and $\gamma$-D posteriors maintain stable and consistent performance. In particular, the DPD posterior tends to outperform the $\gamma$-D posterior in small-sample and high-dimensional settings, although their performance becomes comparable as the sample size increases. The applications to real wind direction data and the gene expression data also illustrated that the proposed methods yield robust posterior summaries that are not distorted by outlying observations.

Overall, the proposed robust generalised Bayesian inference framework offers a flexible, computationally tractable, and theoretically justified approach for directional data analysis under contamination. 
Potential extensions of this work include the development of adaptive selection methods for tuning parameters, applications to more complex directional models such as the Fisher-Bingham distribution \citep{kent1982fisher}, and theoretical investigations of robustness in hierarchical or regression-type models \citep[e.g.][]{holmquist2017two, Scealy2011regression, Scealy2019scaled} for directional data. These directions represent promising avenues for future research.

\section*{Acknowledgement}
This work was supported by JSPS KAKENHI Grant Numbers JP23K13019, JP24K20746, and JP25H00546.

\def\thesection{Appendix}
\def\thesubsection{A.\arabic{subsection}}
\section{}
\def\theequation{A.\arabic{equation}}
\def\thethm{A.\arabic{thm}}
\def\thelem{A.\arabic{lem}}

\subsection{Derivative functions and Expectations}

From $(\dd/\dd x)I_{\nu}(x) = (\nu/x)I_{\nu}(x) + I_{\nu+ 1}(x)$ \citep[see][Appendix]{mardia2000directional}, we obtain
\begin{align}
    \frac{\dd}{\dd x}\log I_{\nu}(x) &= \frac{(\nu/x)I_{\nu}(x) + I_{\nu+ 1}(x)}{I_{\nu}(x)} = \frac{\nu}{x} + \frac{I_{\nu + 1}(x)}{I_{\nu}(x)}, \\
    \frac{\dd}{\dd x}A_{p}(x) &= 1 - A_{p}(x)^2 - \frac{p - 1}{x}A_{p}(x), 
\end{align}
where $A_p(x) = I_{p/2}(x)/I_{(p-2)/2}(x), x \in [0,\infty)$. Consequently, we have
\[
\frac{\partial}{\partial \bm{\xi}} K_p(\bm{\xi}) = \frac{(2 \pi)^{p/2}I_{p/2}(\|\bm{\xi}\|)}{\| \bm{\xi}\|^{(p-2)/2}} \frac{1}{\|\bm{\xi}\|}\bm{\xi}. 
\]
We summarise the gradient (score) functions for each loss. 
The gradient for each loss functions are given by
\begin{align}
\partial\ell_{KL}(\bm{x}; \bm{\xi}) &= -\left\{\bm{x} - \frac{A_p(\|\bm{\xi}\|)}{\|\bm{\xi}\|}\bm{\xi}\right\}, \\
    \partial\ell_{\alpha}(\bm{x}; \bm{\xi}) &= -\frac{\exp(\alpha \bm{x}^{\top}\bm{\xi})}{K_p(\bm{\xi})^{\alpha}}\left\{\bm{x} - \frac{A_p(\|\bm{\xi}\|)}{\|\bm{\xi}\|}\bm{\xi}\right\}\\ 
    &\hspace{10pt} + \frac{K_p((1 + \alpha)\bm{\xi})}{K_p(\bm{\xi})^{\alpha + 1}}\left(A_p((1 + \alpha)\|\bm{\xi}\|) - A_p(\|\bm{\xi}\|)\right)\frac{1}{\|\bm{\xi}\|}\bm{\xi}, \\
\partial\ell_{\gamma}(\bm{x}; \bm{\xi}) &= -\frac{\exp(\gamma \bm{x}^{\top}\bm{\xi})}{K_p((1 + \gamma)\bm{\xi})^{\gamma/(1 + \gamma)}}\left\{\bm{x} - \frac{A_p((1 + \gamma)\|\bm{\xi}\|)}{\|\bm{\xi}\|}\bm{\xi}\right\}. 
\end{align}
Also, we summarise the Hessian functions for each loss function, given by
\begin{align}
&\partial\partial^{\top}\ell_{KL}(\bm{x}; \bm{\xi}) =  \frac{A_p(\|\bm{\xi}\|)}{\|\bm{\xi}\|}\bm{I} + \frac{1}{\|\bm{\xi}\|^2}\left\{1 - A_p(\|\bm{\xi}\|)^2 - \frac{p}{\|\bm{\xi}\|}A_p(\|\bm{\xi}\|)\right\} \bm{\xi}\bm{\xi}^{\top}, \\
&\partial\partial^{\top}\ell_{\alpha}(\bm{x}; \bm{\xi}) = -\frac{\alpha\exp(\alpha \bm{x}^{\top}\bm{\xi})}{K_p(\bm{\xi})^{\alpha}}\left\{\bm{x} - \frac{A_p(\|\bm{\xi}\|)}{\|\bm{\xi}\|}\bm{\xi}\right\}\left\{\bm{x} - \frac{A_p(\|\bm{\xi}\|)}{\|\bm{\xi}\|}\bm{\xi}\right\}^{\top}\\
&\hspace{10pt}+ \frac{\exp(\alpha\bm{x}^{\top}\bm{\xi})}{K_p(\bm{\xi})^{\alpha}}\left\{ \frac{A_p(\|\bm{\xi}\|)}{\|\bm{\xi}\|}\bm{I} + \frac{1}{\|\bm{\xi}\|^2}\left\{1 - A_p(\|\bm{\xi}\|)^2 - \frac{p}{\|\bm{\xi}\|}A_p(\|\bm{\xi}\|)\right\} \bm{\xi}\bm{\xi}^{\top} \right\}\\
&\hspace{10pt} + (\alpha + 1)\frac{K_p((1 + \alpha)\bm{\xi})}{K_p(\bm{\xi})^{\alpha + 1}}\left(A_p((1 + \alpha)\|\bm{\xi}\|) - A_p(\|\bm{\xi}\|)\right)^2\frac{1}{\|\bm{\xi}\|^2}\bm{\xi}\bm{\xi}^{\top}\\
&\hspace{10pt}  +\frac{K_p((1 + \alpha)\bm{\xi})}{K_p(\bm{\xi})^{\alpha + 1}}\left[ \frac{A_p(\|(1 + \alpha)\bm{\xi}\|)}{\|\bm{\xi}\|}\bm{I} \right. \\
&\hspace{10pt} \left. \hspace{30pt} + \frac{(1 + \alpha)}{\|\bm{\xi}\|^2}\left\{1 - A_p((1 + \alpha)\|\bm{\xi}\|)^2 - \frac{p}{\|(1 + \alpha)\bm{\xi}\|}A_p(\|(1 + \alpha)\bm{\xi}\|)\right\} \bm{\xi}\bm{\xi}^{\top} \right]\\
&\hspace{10pt}  - \frac{K_p((1 + \alpha)\bm{\xi})}{K_p(\bm{\xi})^{\alpha + 1}}\left\{ \frac{A_p(\|\bm{\xi}\|)}{\|\bm{\xi}\|}\bm{I} + \frac{1}{\|\bm{\xi}\|^2}\left\{1 - A_p(\|\bm{\xi}\|)^2 - \frac{p}{\|\bm{\xi}\|}A_p(\|\bm{\xi}\|)\right\} \bm{\xi}\bm{\xi}^{\top} \right\}, \\
&\partial\partial^{\top}\ell_{\gamma}(\bm{x}; \bm{\xi}) = \frac{-\gamma\exp(\gamma \bm{x}^{\top}\bm{\xi})}{K_p((1 + \gamma)\bm{\xi})^{\gamma/(1 + \gamma)}}\left\{\bm{x} - \frac{A_p((1 + \gamma)\|\bm{\xi}\|)}{\|\bm{\xi}\|}\bm{\xi}\right\}\left\{\bm{x} - \frac{A_p((1 + \gamma)\|\bm{\xi}\|)}{\|\bm{\xi}\|}\bm{\xi}\right\}^{\top}\\
&\hspace{10pt}+ \frac{\exp(\gamma \bm{x}^{\top}\bm{\xi})}{K_p((1 + \gamma)\bm{\xi})^{\gamma/(1 + \gamma)}}\left\{ \frac{A_p((1 + \gamma)\|\bm{\xi}\|)}{\|\bm{\xi}\|}\bm{I}\right.\\ 
&\hspace{10pt}\left. + \frac{1 + \gamma}{\|\bm{\xi}\|^2}\left\{1 - A_p((1 + \gamma)\|\bm{\xi}\|)^2 - \frac{p}{(1 + \gamma)\|\bm{\xi}\|}A_p((1 + \gamma)\|\bm{\xi}\|)\right\} \bm{\xi}\bm{\xi}^{\top} \right\}. 
\end{align} 

The first integration can be calculated by using the fact that if $ X \sim \textrm{vM}_p(\bm{\xi})$, then $\E(X) = A_p(\|\bm{\xi}\|)\bm{\xi}/\|\bm{\xi}\|$. \citep[See, for example][pp.169]{mardia2000directional}. From this, it immediately follows that
\begin{align} 
    \int_{\mcal{S}_{p}} \bm{y} \exp\{(1 + \beta) \bm{\xi}^{\top}\bm{y}\} \dd \bm{y} = K_p((1 + \beta)\bm{\xi}) \frac{A_p((1 + \beta)\|\bm{\xi}\|)}{\|\bm{\xi}\|}\bm{\xi}. 
\end{align}
Next, we have 
\begin{align}
    &\int_{S_p} f_{\bm{\xi}}(\bm{y})\exp(\beta \bm{y}^{\top}\bm{\xi})\left\{\bm{y} - \frac{A_p(\|\bm{\xi}\|)}{\|\bm{\xi}\|}\bm{\xi}\right\}\left\{\bm{y} - \frac{A_p(\|\bm{\xi}\|)}{\|\bm{\xi}\|}\bm{\xi}\right\}^{\top} \dd \bm{y} \\
    & = \frac{K_p((\beta + 1) \bm{\xi})}{K_p(\bm{\xi})}\int_{S_p} f_{(\beta + 1)\bm{\xi}}(\bm{y})\left\{\bm{y} - \frac{A_p(\|\bm{\xi}\|)}{\|\bm{\xi}\|}\bm{\xi}\right\}\left\{\bm{y} - \frac{A_p(\|\bm{\xi}\|)}{\|\bm{\xi}\|}\bm{\xi}\right\}^{\top} \dd \bm{y} \\
    & = \frac{K_p((\beta + 1) \bm{\xi})}{K_p(\bm{\xi})}\left[\int_{S_p} f_{(\beta + 1)\bm{\xi}}(\bm{y})\left\{\bm{y} - \frac{A_p((\beta + 1)\|\bm{\xi}\|)}{\|\bm{\xi}\|}\bm{\xi}\right\}\left\{\bm{y} - \frac{A_p((\beta + 1)\|\bm{\xi}\|)}{\|\bm{\xi}\|}\bm{\xi}\right\}^{\top} \dd \bm{y}\right.\\
    & \left. + \frac{(A_p((\beta + 1)\|\bm{\xi}\|) - A_p(\|\bm{\xi}\|))^2}{\|\bm{\xi}\|^2}\bm{\xi}\bm{\xi}^{\top}\right]\\
    & = \frac{K_p((\beta + 1) \bm{\xi})}{K_p(\bm{\xi})}\left[\frac{A_p((\beta + 1)\|\bm{\xi}\|)}{\|(\beta + 1)\bm{\xi}\|}\bm{I}\right.\\
    & \hspace{20pt} + \frac{1}{\|\bm{\xi}\|^2}\left\{1 - A_p((\beta + 1)\|\bm{\xi}\|)^2 - \frac{p}{\|(\beta + 1)\bm{\xi}\|}A_p((\beta + 1)\|\bm{\xi}\|)\right\} \bm{\xi}\bm{\xi}^{\top}\\
     & \left. \hspace{20pt} + \frac{(A_p((\beta + 1)\|\bm{\xi}\|) - A_p(\|\bm{\xi}\|))^2}{\|\bm{\xi}\|^2}\bm{\xi}\bm{\xi}^{\top}\right]. 
\end{align}

\subsection{Information Matrices}
\label{app:information}

In the case of the ordinary log-likelihood, we have 
\begin{align}
    \bm{I}^{(KL)}(\bm{\xi}) &= \bm{J}^{(KL)}(\bm{\xi})
    = \E\left[\partial\ell_{KL}(\bm{X}; \bm{\xi})\partial^{\top}\ell_{KL}(\bm{X}; \bm{\xi})\right] 
    = \E\left[\partial\partial^{\top}\ell_{KL}(\bm{X}; \bm{\xi})\right]\\
    & =  \frac{A_p(\|\bm{\xi}\|)}{\|\bm{\xi}\|}\bm{I} + \frac{1}{\|\bm{\xi}\|^2}\left\{1 - A_p(\|\bm{\xi}\|)^2 - \frac{p}{\|\bm{\xi}\|}A_p(\|\bm{\xi}\|)\right\} \bm{\xi}\bm{\xi}^{\top}\\
    & = a_1^{(KL)}\bm{I} + a_2^{(KL)}\frac{1}{\|\bm{\xi}\|^2}\bm{\xi}\bm{\xi}^{\top}.
\end{align}
In the case of the density power cross entropy, we have 
\begin{align}
    &\bm{I}^{(\alpha)}(\bm{\xi}) = \E\left[\partial\ell_{\alpha}(\bm{X}; \bm{\xi})\partial^{\top}\ell_{\alpha}(\bm{X}; \bm{\xi})\right]
      = \frac{K_p((2\alpha + 1) \bm{\xi})}{K_p(\bm{\xi})^{2\alpha + 1}}\left(a_1^{(\alpha)}\bm{I} + a_2^{(\alpha)}\frac{1}{\|\bm{\xi}\|^2}\bm{\xi}\bm{\xi}^{\top}\right),  \\
&\bm{J}^{(\alpha)}(\bm{\xi}) = \E\left[\partial\partial^{\top}\ell_{\alpha}(\bm{X}; \bm{\xi})\right]
      = \frac{K_p((\alpha + 1) \bm{\xi})}{K_p(\bm{\xi})^{\alpha + 1}}\left(b_1^{(\alpha)}\bm{I} + b_2^{(\alpha)}\frac{1}{\|\bm{\xi}\|^2}\bm{\xi}\bm{\xi}^{\top}\right),  
\end{align}
where 
\begin{align}
a_1^{(\alpha)} &= \frac{A_p((2\alpha + 1)\|\bm{\xi}\|)}{\|(2\alpha + 1)\bm{\xi}\|}, \\
a_2^{(\alpha)} &= \left\{1 - A_p((2\alpha + 1)\|\bm{\xi}\|)^2 - \frac{p}{\|(2\alpha + 1)\bm{\xi}\|}A_p((2\alpha + 1)\|\bm{\xi}\|)\right\} \\
& + (A_p((2\alpha + 1)\|\bm{\xi}\|) - A_p(\|\bm{\xi}\|))^2\\
& - \frac{K_p((1 + \alpha)\bm{\xi})}{K_p(\bm{\xi})}\left(A_p((1 + \alpha)\|\bm{\xi}\|) - A_p(\|\bm{\xi}\|)\right)^2, \\
b_1^{(\alpha)} &= \frac{A_p((\alpha + 1)\|\bm{\xi}\|)}{\|(\alpha + 1)\bm{\xi}\|}, \\
b_2^{(\alpha)} & = \left\{1 - A_p((\alpha + 1)\|\bm{\xi}\|)^2 - \frac{p}{\|(\alpha + 1)\bm{\xi}\|}A_p((\alpha + 1)\|\bm{\xi}\|)\right\}\\
& + \left(A_p((1 + \alpha)\|\bm{\xi}\|) - A_p(\|\bm{\xi}\|)\right)^2. 
\end{align}
In the case of the $\gamma$-cross entropy, we have 
\begin{align}
 &\bm{I}^{(\gamma)}(\bm{\xi}) = \E\left[\partial\ell_{\gamma}(\bm{X}; \bm{\xi})\partial^{\top}\ell_{\gamma}(\bm{X}; \bm{\xi})\right]\\
     & = \frac{K_p((2\gamma + 1) \bm{\xi})}{K_p(\bm{\xi})K_p((1 + \gamma)\bm{\xi})^{2\gamma/(1 + \gamma)}}\left(a_1^{(\gamma)}\bm{I} + a_2^{(\gamma)}\frac{1}{\|\bm{\xi}\|^2}\bm{\xi}\bm{\xi}^{\top}\right),  \\
&\bm{J}^{(\gamma)}(\bm{\xi}) = \E\left[\partial\partial^{\top}\ell_{\gamma}(\bm{X}; \bm{\xi})\right] \\
& = \frac{K_p((\gamma + 1) \bm{\xi})^{1/(1 + \gamma)}}{K_p(\bm{\xi})}\left(b_1^{(\gamma)}\bm{I} + b_2^{(\gamma)}\frac{1}{\|\bm{\xi}\|^2}\bm{\xi}\bm{\xi}^{\top}\right). 
\end{align}

where 
\begin{align}
a_1^{(\gamma)} &= \frac{A_p((2\gamma + 1)\|\bm{\xi}\|)}{\|(2\gamma + 1)\bm{\xi}\|}, \\
a_2^{(\gamma)} &= \left\{1 - A_p((2\gamma + 1)\|\bm{\xi}\|)^2 - \frac{p}{\|(2\gamma + 1)\bm{\xi}\|}A_p((2\gamma + 1)\|\bm{\xi}\|)\right\} \\
& + (A_p((2\gamma + 1)\|\bm{\xi}\|) - A_p((1 + \gamma)\|\bm{\xi}\|))^2, \\
b_1^{(\gamma)} &= \frac{A_p((\gamma + 1)\|\bm{\xi}\|)}{\|(\gamma + 1)\bm{\xi}\|}, \\
b_2^{(\gamma)} & = \left\{1 - A_p((\gamma + 1)\|\bm{\xi}\|)^2 - \frac{p}{\|(\gamma + 1)\bm{\xi}\|}A_p((\gamma + 1)\|\bm{\xi}\|)\right\}. 
\end{align}

\subsection{Selection of tuning parameters}
\label{app:Select}

The selection of the tuning parameters $\alpha$ and $\gamma$ is highly challenging. 
Although various methods have been proposed, for example, \cite{Basak2021optimal, sugasawa2021selection,Warwick2005choosing, yonekura2023adaptation}, none offers a universally appropriate or theoretically optimal choice. 
The tuning parameters $\alpha$ and $\gamma$ control the degree
of robustness, that is, if we set large $\alpha$ and $\gamma$, we obtain higher robustness.
However, there is a trade-off between the robustness and efficiency of estimators.
One of the solutions for this problem is to use the asymptotic relative efficiency (ARE) \citep[see, for example, ][so on.]{ghosh2016robust, nakagawa2021default}. 

\begin{figure}
\begin{minipage}[t]{.45\textwidth}
\centering
\includegraphics[width=6cm]{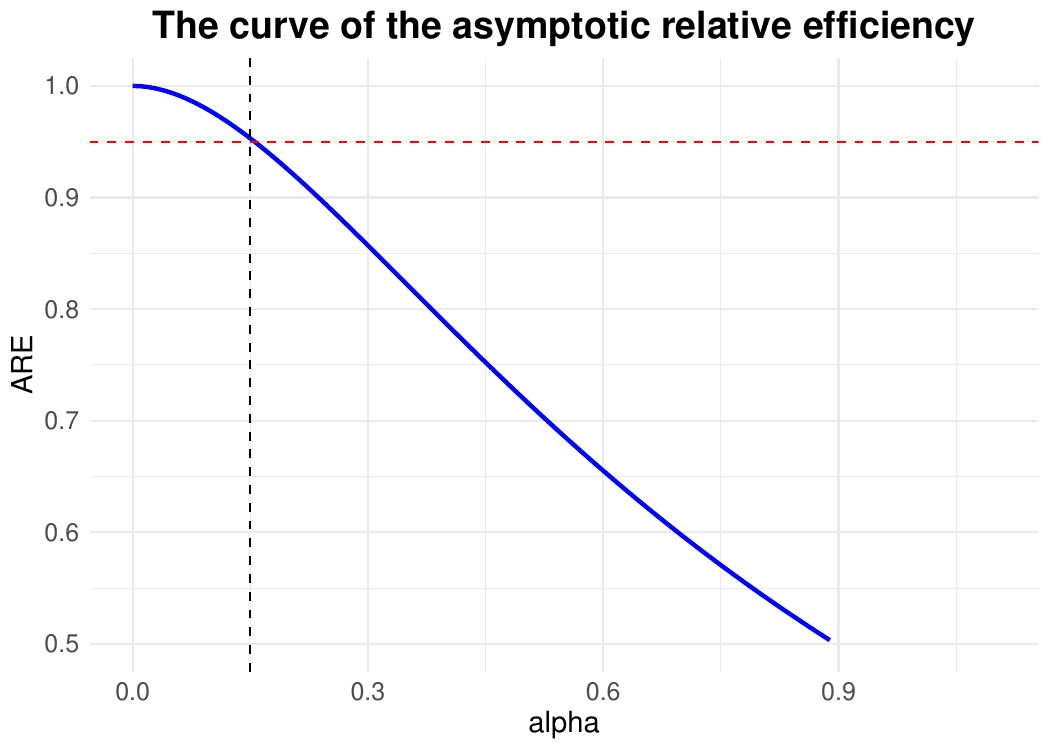}
\end{minipage}
\begin{minipage}[t]{.45\textwidth}
\centering
\includegraphics[width= 6cm]{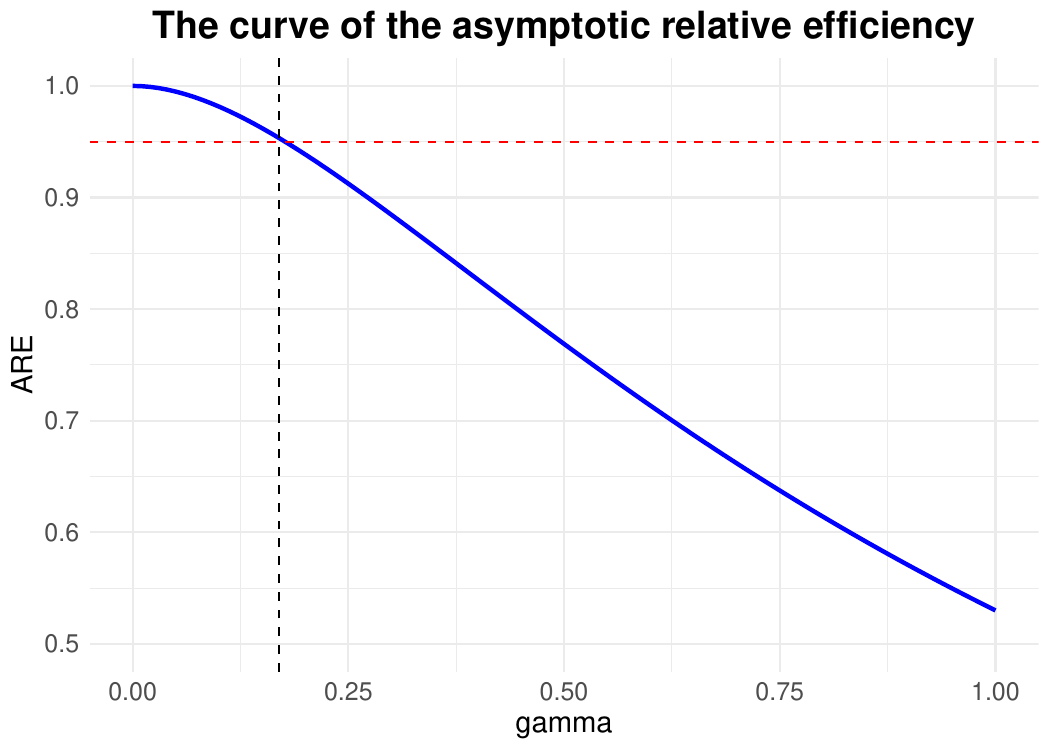}
\end{minipage}
\caption{The figures display curve of the asymptotic relative efficiency for the density power-posterior and $\gamma$-D posterior means in the case of dimension $p = 2$. The dashed red line indicates $95\%$, and the dashed black line indicates the hyperparameter ($\alpha$, $\gamma$) values corresponding to the 95th percentile. }
\label{app:fig_ARE}
\end{figure}

In general, the asymptotic relative efficiency of the proposed posterior mean $\tilde{\bm{\xi}}^{(d)}$ of $p$-dimensional parameter $\bm{\xi}$ relative to the ordinary posterior mean $\tilde{\bm{\xi}}$ is defined by
\begin{align*}
\mathrm{ARE}(\tilde{\bm{\xi}}^{(d)},\tilde{\bm{\xi}}):=\left(\frac{\det\left(\bm{V}(\bm{\xi})\right)}{\det\left(\bm{V}^{(d)}(\bm{\xi})\right)}\right)^{1/p}, 
\end{align*}
where $\bm{V}(\bm{\xi})$ and $\bm{V}^{(d)}(\bm{\xi})$ are the asymptotic covariance matrices of $\tilde{\bm{\xi}}$ and $\tilde{\bm{\xi}}^{(d)}$ \citep[see, for example, ][]{serfling1980book}.
 We now calculate the $\mathrm{ARE}(\tilde{\bm{\xi}}^{(\alpha)},\tilde{\bm{\xi}})$ and $\mathrm{ARE}(\tilde{\bm{\xi}}^{(\gamma)},\tilde{\bm{\xi}})$ in our simulation setting where $\tilde{\bm{\xi}}^{(\alpha)}$ and $\tilde{\bm{\xi}}^{(\gamma)}$ are the DPD posterior and $\gamma$-D posterior means, respectively. 
After some calculations, the asymptotic relative efficiency is given by
\begin{align*}
h_1(\alpha) &\coloneqq \mathrm{ARE}(\hat{\bm{\xi}}^{(\alpha)},\hat{\bm{\xi}}) =\left(\frac{\det\left(\bm{I}^{(KL)}(\bm{\xi})^{-1}\right)}{\det\left(\bm{J}^{(\alpha)}(\bm{\xi})^{-1}\bm{I}^{(\alpha)}(\bm{\xi})\bm{J}^{(\alpha)}(\bm{\xi})^{-1}\right)}\right)^{1/p}\\
& = \frac{K_p((\alpha + 1) \bm{\xi})^2\left(b_1^{(\alpha)} + b_2^{(\alpha)}\right)^{2/p}\{b_1^{(\alpha)}\}^{2(1-1/p)}}{K_p( \bm{\xi})K_p((2\alpha + 1) \bm{\xi})\left\{\left(a_1^{(\alpha)} + a_2^{(\alpha)}\right)\left(a_1^{(KL)} + a_2^{(KL)}\right)\right\}^{1/p}\{a_1^{(\alpha)}a_1^{(KL)}\}^{1-1/p}}, \\
h_2(\gamma) &\coloneqq \mathrm{ARE}(\hat{\bm{\xi}}^{(\gamma)},\hat{\bm{\xi}}) =\left(\frac{\det\left(\bm{I}^{(KL)}(\bm{\xi})^{-1}\right)}{\det\left(\bm{J}^{(\gamma)}(\bm{\xi})^{-1}\bm{I}^{(\gamma)}(\bm{\xi})\bm{J}^{(\gamma)}(\bm{\xi})^{-1}\right)}\right)^{1/p}\\
& = \frac{K_p((\gamma + 1) \bm{\xi})^{4\gamma/(1 + \gamma)}\left(b_1^{(\gamma)} + b_2^{(\gamma)}\right)^{2/p}\{b_1^{(\gamma)}\}^{2(1-1/p)}}{K_p( \bm{\xi})K_p((2\gamma + 1) \bm{\xi})\left\{\left(a_1^{(\gamma)} + a_2^{(\gamma)}\right)\left(a_1^{(KL)} + a_2^{(KL)}\right)\right\}^{1/p}\{a_1^{(\gamma)}a_1^{(KL)}\}^{1-1/p}}, \\
\end{align*}
for $\alpha, \gamma>0$. 
For example, if we require the value of the asymptotic relative efficiency $\mathrm{ARE}=0.95$, we may choose the values of $\alpha$ and $\gamma$ as the solution of the equation $h_1(\alpha) = h_2(\gamma) =0.95$.
Figure \ref{app:fig_ARE} presents the curves of functions $h_1(\alpha)$ and $ h_2(\gamma)$ in the case of dimension $p = 2$, where the dashed black lines denote the hyperparameter values corresponding to the 95th percentile.

\bibliographystyle{apalike} 
\bibliography{References.bib}

\end{document}